\documentclass[twocolumn,amsmath,amssymb,aps, prb]{revtex4-2}
\usepackage{graphicx}
\usepackage{color,soul}
\usepackage{amsmath,amssymb}
\usepackage{mathrsfs}
\usepackage{color}
\usepackage{afterpage}
\usepackage{pifont}
\usepackage[version=3]{mhchem}
\usepackage{natbib}
\usepackage{soul}
\usepackage[caption=false]{subfig}
\usepackage{xtab,afterpage}
\usepackage{lipsum}
\usepackage{array}
\usepackage{multirow}
\usepackage{bm}
 
\usepackage[colorlinks, citecolor=blue,urlcolor=blue, linkcolor=blue, bookmarks=false]{hyperref}
\hypersetup{colorlinks=true , citecolor=blue, urlcolor=blue, linkcolor=blue}
\begin{document}

\title{Spontaneous Anomalous Hall Effect in Two-Dimensional Altermagnets}

\author{Sajjan Sheoran} 
\author{Pratibha Dev}
\affiliation{Department of Physics and Astronomy, Howard University, Washington D.C., USA}


\begin{abstract}


The anomalous Hall effect (AHE) is an efficient tool for detecting the N\'eel vector in collinear compensated magnets with spin-split bands, known as altermagnets (AMs). Here, we establish design principles for obtaining non-zero anomalous Hall conductivity in the recently proposed two-dimensional (2D) AMs using spin and magnetic group symmetry analysis. We show that only two of the seven nontrivial spin layer groups exhibit an unconventional in-plane AHE in which the N\'eel vector lies within the plane of the Hall current. Through first-principles simulations on bilayers of MnPSe$_3$ and MnSe, we demonstrate the validity of our group theoretic framework for obtaining AHE with $d$ and $i$-wave altermagnetic orders, depending on the stacking of the bilayers. We find that the spin group symmetry is successful in determining the linear and cubic dependence of anomalous Hall conductivity in N\'eel vector space, although AHE is a relativistic effect. This work shows that the AHE in 2D AMs can probe the altermagnetic order and N\'eel vector reversal, thereby facilitating the miniaturization of altermagnetic spintronics.

\end{abstract}
\maketitle

\section{Introduction} 
Altermagnetism (AM) has emerged as a novel class of collinear magnetism characterized by time-reversal symmetry ($T$) breaking in momentum space along with compensated magnetic order in real space~\cite{yuan2021prediction, yuan2021strong, vsmejkal2022beyond, vsmejkal2022emerging, hayami2020bottom}.  Unlike conventional antiferromagnets (AFMs), where the opposite spin sublattices are connected by the inversion ($P$) and/or translation ($\tau$) operations, in AMs they are connected by mirror-rotation symmetries. Hence, AMs show eV-scale nonrelativistic spin splittings~\cite{vsmejkal2022beyond}, leading to highly spin-polarized currents, which are characteristic of ferromagnets (FMs). Since there is no net magnetization, AMs also show ultrafast switching dynamics and resilience towards stray fields, similar to that shown by AFMs~\cite{vzutic2004spintronics, tsymbal2019spintronics, nvemec2018antiferromagnetic,vzelezny2018spin}.  This combination of properties of both FMs and AFMs in a single material makes AMs interesting not only  for fundamental research, but also for spintronics-based applications by allowing for facile control and detection of spin states using different means, including electric or optical fields. Broken $T$-symmetry effects in AMs are experimentally detected through angle-resolved photoemission spectroscopy~\cite{lee2024broken, krempasky2024altermagnetic}, spin-to-charge interconversion~\cite{bai2023efficient, liao2024separation}, and the anomalous Hall effect (AHE)~\cite{vsmejkal2022anomalous}.  To date, the experimental observation of altermagnetic effects is still limited to three-dimensional (3D) materials, i.e., MnTe~\cite{gonzalez2023spontaneous}, Mn$_5$Si$_3$~\cite{reichlova2024observation}, CrSb~\cite{reimers2024direct}, and RuO$_2$~\cite{bai2023efficient}.

\begin{table*}[t]
	\caption{Symmetry constraints on the components of the anomalous Hall conductivity tensor imposed by the magnetic point group operations. \ding{51} and \ding{55} denotes symmetry allowed and forbidden components, respectively. The $z$ ($x$) denotes the out-of-plane (in-plane) direction.}
	\centering
	\label{t1}
	\begin{tabular*}{\textwidth}{@{\extracolsep{\fill}}l c c c c c c c c c c c c c c c}
		\hline
		\hline
		& $P$ & $C_n^z$ & $C_n^x$ & $M_z$ & $M_x$ & $S_{4,6}^z$ & $S_{4,6}^x$ & $T$ & $PT$ & $TC_2^z$ & $TC_2^x$ & $TM_z$ & $TM_x$ & $TC_{3,4,6}^{x,z}$ & $TS_{4,6}^{x,z}$ \\
		\hline
		$\sigma_{xy}$ & \ding{51}  & \ding{51}  & \ding{55} &\ding{51} & \ding{55} & \ding{51} &\ding{55} & \ding{55} &\ding{55} & \ding{55} & \ding{51} & \ding{55} & \ding{51} & \ding{55} & \ding{55}\\
		$\sigma_{xz}$ & \ding{51}  & \ding{55} & \ding{55} & \ding{55} & \ding{55} &  \ding{55} &  \ding{55} & \ding{55} & \ding{55} & \ding{51} & \ding{51} & \ding{51} & \ding{51} & \ding{55} &\ding{55} \\
		$\sigma_{yz}$ & \ding{51}  & \ding{55} & \ding{51} & \ding{55} & \ding{51} &  \ding{55} & \ding{51} & \ding{55} &\ding{55} & \ding{51} & \ding{55} & \ding{51} & \ding{55} & \ding{55}&\ding{55} \\
		\hline
		\hline
	\end{tabular*}
\end{table*}

The discovery of magnetic ordering in atomically thin materials opens up new possibilities for miniaturizing devices to the two-dimensional (2D) limit. However, achieving AM in 2D systems is difficult due to additional symmetry constraints. In 2D layers, the electronic bands are dispersionless along the out-of-plane direction~\cite{pan2024general, zeng2024description, zeng2024bilayer, he2023nonrelativistic}. As a result, in the 2D limit, the two-fold rotation along the $z$-axis ($C_{2}^{z}$) and horizontal mirror ($M_z$) symmetry transform $\boldsymbol{k} = (k_x, k_y)$ like $P$ and $\tau$, respectively. Therefore, for altermagnetism to emerge, the opposite spin sublattices must not be connected by \( M_z \) and/or \( C_{2}^{z} \)~\cite{zeng2024description}. In spite of these exacting requirements, AM in 2D materials was predicted theoretically using a high-throughput computational approach (RuF$_4$, FeBr$_3$~\cite{sodequist2024two, milivojevic2024interplay}), bilayer stackings (CrSBr, MnBi$_2$Te$_4$~\cite{pan2024general, zeng2024bilayer}) and twisting bilayers~\cite{he2023nonrelativistic, liu2024twisted, sheoran2024nonrelativistic}.   The nonrelativistic understanding of 2D altermagnetism was recently established in Refs.~\cite{pan2024general, zeng2024description, zeng2024bilayer, he2023nonrelativistic, he2023nonrelativistic, liu2024twisted, sheoran2024nonrelativistic, sodequist2024two, milivojevic2024interplay}.
However, unlike 3D bulk AMs, a comprehensive analysis of the magnetotransport effects (such as anomalous Hall and anomalous Nernst effects), and novel N\'eel vector detection methods for 2D AMs is still lacking. 

Our work based on symmetry analysis and first- principles simulations elucidates the stringent symmetry requirement for observing AHE in the 2D AMs. We show an unconventional periodic dependence of AHE on the N\'eel-vector space, where the N\'eel vector lies in the plane of the Hall current. This is in contrast to the conventional FM-like Hall response, where Hall current is perpendicular to the magnetization~\cite{onoda2006intrinsic, shindou2001orbital}. By performing density functional theory (DFT) simulations on  bilayers of MnPSe$_3$ and MnSe, which are  $PT$-symmetric as monolayers, we achieved $d$- and $i$-wave altermagnetism with an in-plane anomalous Hall response as large as for the well-known bulk AMs\textemdash MnTe~\cite{gonzalez2023spontaneous, kluczyk2024coexistence} and Mn$_5$Si$_3$~\cite{reichlova2024observation,leiviska2024anisotropy}. Furthermore, we reveal the unique relationship of the AHE with the spin group symmetries and nonrelativsitic spin-degenerate nodal lines. Moreover, the in-plane AHE is shown to be an efficient tool to probe the altermagnetic order and 180$^\circ$ N\'eel vector reversal in 2D AMs.  

\section{Calculation Methods} 
DFT calculations have been carried out using the projector-augmented wave method~\cite{kresse1999ultrasoft}, as implemented in the VASP package~\cite{kresse1996efficient}. The Perdew-Burke-Ernzerhof (PBE)~\cite{perdew1996generalized} functional within the generalized-gradient approximation, along with a Hubbard U correction, are employed to accurately describe electronic interactions. Following Refs.~\cite{liu2023tunable, rybak2024magneto}, an effective U value of 3.0 eV was applied to the Mn-$d$ orbitals using the approach of Dudarev \textit{et al}.~\cite{dudarev1998electron}. The Grimme-D3 scheme was used to account for the vdW interactions~\cite{grimme2006semiempirical}. 

 The AHE can be described by an order parameter called the Hall vector ($\boldsymbol{\sigma}_\textrm{H}$), with the anomalous Hall current density given by $\boldsymbol{J}=\boldsymbol{\sigma}_\textrm{H} \times \boldsymbol{E} $, where $\boldsymbol{E}$ is the electric field.  The Hall vector is defined as: $\boldsymbol{\sigma}_\textrm{H}$=($\sigma_{zy}, \sigma_{xz}, \sigma_{yx})$~\cite{nagaosa2010anomalous}, with $\sigma_{\alpha\beta}$ for $\alpha \neq \beta$ being the antisymmetric components of the conductivity tensor. At the microscopic level, the intrinsic contribution to anomalous Hall conductivity (AHC) is expressed as 
\begin{equation}
	\sigma_{\mathrm{\alpha\beta}}=- \epsilon_{\alpha \beta \gamma} \frac{e^2}{\hbar} \int \frac{d \boldsymbol{k}}{(2 \pi)^3} \sum_n f({\mathcal{E}}_{n, \boldsymbol{k}}) \Omega_{\gamma}(n, \boldsymbol{k})
	\label{E1}
\end{equation}
where $\epsilon_{\alpha \beta \gamma}$ is Levi-Civita symbol. $\mathcal{E}_{n, \boldsymbol{k}}$ and $\Omega_{\gamma}$ are the energy eigenvalue and Berry curvature, respectively, of the band with quantum number $n$. $f(\mathcal{E}_{n, \boldsymbol{k}})$ is the Fermi-Dirac distribution function. At zero temperature, the summation in Eq.~\ref{E1} reduces to a sum over occupied bands. To simulate AHC within DFT, we constructed a tight-binding Hamiltonian from maximally localized Wannier functions, which were obtained through Wannier90~\cite{mostofi2008wannier90}. Wannier interpolation of  Berry curvature and AHC were postprocessed using the ``\textit{pruned FFT}"-based WannierBerri~\cite{tsirkin2021high} package. Note that rapid variations of Berry curvature are taken into account by integrating the Brillouin Zone on a dense $k$-mesh of 125$\times$125$\times$1 and with recursive adaptive mesh refinement. We conducted symmetry analysis using Ref.~\cite{dresselhaus2007group}, FINDSYM~\cite{stokes2005findsym}, Bilbao Crystallographic Server~\cite{aroyo2006bilbao}, MAGNDATA~\cite{gallego2016magndata}, and AMCHECK~\cite{smolyanyuk2024codebase}.

\begin{figure}[htp]
	\includegraphics[width=8.6cm]{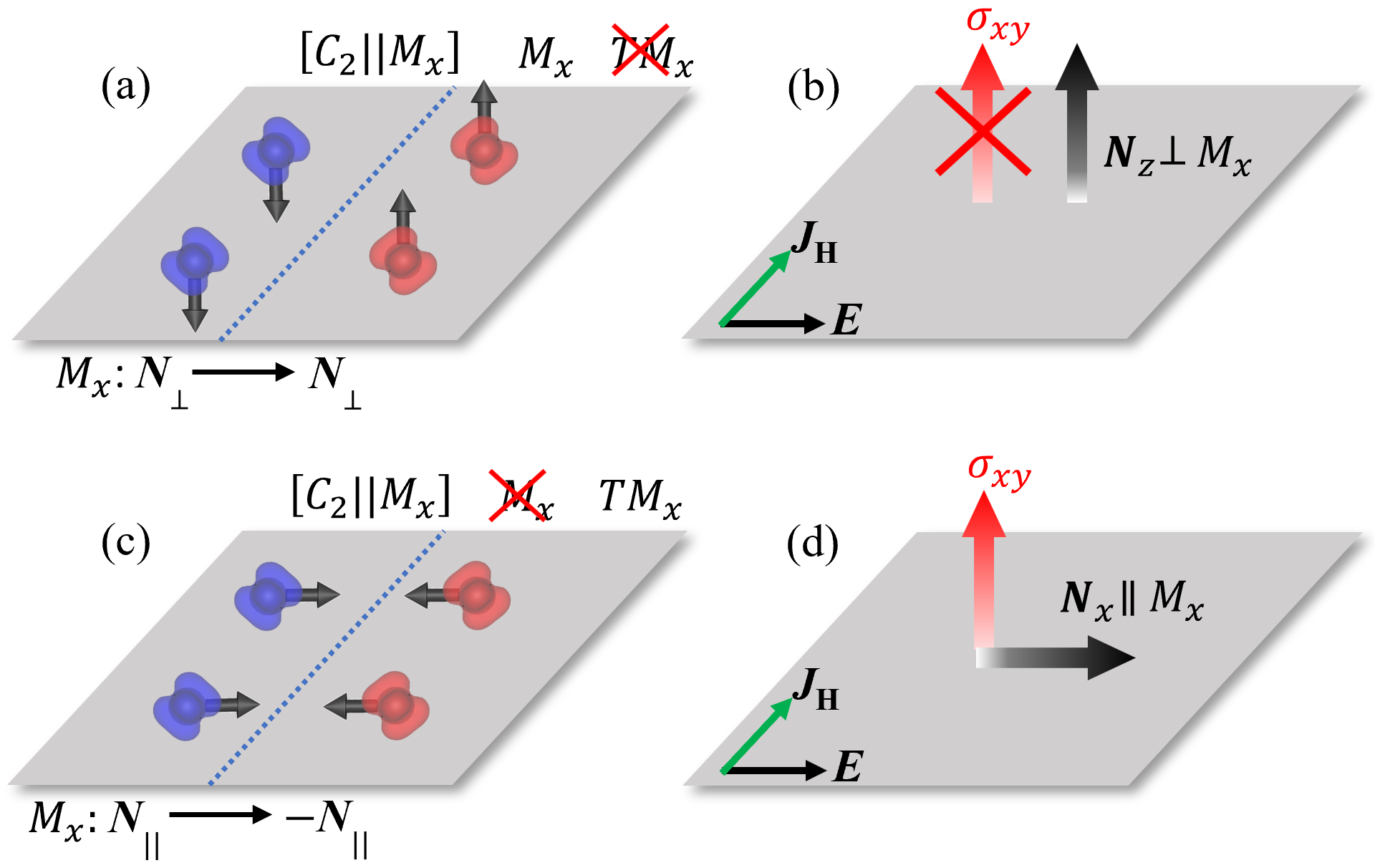}
	\caption{Schematic diagram showing how symmetries determine the presence or absence of AHE in 2D altermagnets. In this example, the spin-group symmetry [$C_2||M_x$] connects the opposite spin-sublattices. (a) $M_x$ symmetry is satisfied when the N\'eel vector ($\boldsymbol{N}$) is perpendicular to the direction of the mirror plane. Note that the direction of mirror plane refers to the direction normal to the mirror (here, $x$-direction). (b) $M_x$ symmetry forbids the AHE (see Table~\ref{t1}). (c) $\boldsymbol{N}$ being parallel to the $x$-direction violates the $M_x$ symmetry and (d) thus the AHE is symmetry allowed.  Since $\boldsymbol{N}$, $\boldsymbol{J}_\textrm{H}$, and $\boldsymbol{E}$ all lie in the same plane (here, the $x$-$y$ plane), this is referred to as the in-plane AHE.}
	\label{p1}
\end{figure}
\section{Results and discussion}
\subsection{Symmetry analysis for AHE in 2D AMs}
Since AMs display spin-splitting of electronic structure without spin-orbit coupling (SOC),  they are best characterized by spin group symmetries [$R_1||R_2$], where $R_1$ and $R_2$ symmetry operations act on the decoupled spin and real space, respectively~\cite{litvin1974spin, vsmejkal2022beyond, jiang2024enumeration, liu2022spin}. Just like in the case of AFMs, symmetry dictates a net zero magnetization in AMs. However, in AFMs, the spin-up and -down bands are degenerate. It is instructive to see which symmetries (in terms of spin group formalism) would result in a traditional AFM versus an AM. Collinear magnets, including AMs, always have spin-only symmetry [$\overline{C}_2||T$], where $\overline{C}_2$ is a two-fold rotation around the axis perpendicular to the collinear spins, followed by  inversion in spin-space~\cite{litvin1974spin}. The [$\overline{C}_2||T$] transforms energy eigenstates as $[\overline{C}_2||T$]$E(s, \boldsymbol{k})$=$E(s, -\boldsymbol{k})$, leading to even-parity spin splitting. The  $[\overline{C}_2||T$][$C_2||P$] and [$C_2||\tau$] transform the energy eigenstates from $E(s, \boldsymbol{k})$ to $E(-s, \boldsymbol{k})$, which lead to spin degeneracy at an arbitrary $k$-point. The $C_2$ operation is the two-fold rotation about an axis perpendicular to collinear spins in spin space and, for simplicity, can be interpreted as spin space inversion. Additionally, for the 2D case ($k_z=0$), the spin-group symmetries \( [C_2 \| M_z] \) and \( [C_2 \| C_{2}^{z}] \) also lead to spin degeneracy (see Sec. I of Supplemental Material (SM)~\cite{SuMa}). Overall, to obtain altermagnetism in 2D materials, opposite spin sublattices should not be connected by crystallographic $P$, $\tau$, $C_{2}^{z}$, and $M_z$. However, there should exist at least one crystallographic symmetry connecting opposite spin sublattices to have symmetry-enforced net zero magnetization. The possible symmetry options are in-plane two-fold rotation (i.e., $[C_2||C_{2}^{x}]$), vertical mirror plane (i.e., $[C_2||M_{x}]$), and out-of-plane four-fold rotation ($[C_2||C_{4}^{z}]$). The 2D AM case with  $[C_2||M_{x}]$ symmetry is highlighted in Fig.~\ref{p1}. 

The symmetry analysis for the relativistic AHE requires the considerations of magnetic symmetry operations acting on coupled spin and real spaces. From the symmetry perspective, the Hall vector, $\boldsymbol{\sigma}_\textrm{H}$, transforms like a pseudovector, similar to magnetization. Therefore, the magnetic symmetry operations impose certain constraints on allowed components, $\sigma_{ij}$, of the system (see Table~\ref{t1} and Sec. II of SM~\cite{SuMa}). In particular, all components of the AHC tensor are symmetry forbidden if a material (2D or 3D) possesses either of the $T$, $PT$, $TC_{3,4,6}^{x,z}$, or $TS_{4,6}^{x,z}$ symmetries. Further constraints determining non-zero AHC are imposed by the 2D nature of the 2D AMs.  As the Hall current is restricted to the plane of the 2D materials, the only component that is experimentally relevant is $\sigma_{xy}$ (taking $z$ as the out-of-plane axis). Hence, the presence of $C_n^x$, $M_x$, $TM_z$, and $TC_2^z$ symmetries suppresses AHE in 2D materials as these symmetries result in zero $\sigma_{xy}$ [see Table~\ref{t1}]. Overall, our symmetry analysis shows that three of the following conditions must be simultaneously satisfied to obtain a 2D AM with AHE: (i) absence of [$C_2||P$], [$C_2||\tau$], [$C_2||M_z$], and [$C_2||C_{2}^{z}$] spin group symmetries that will otherwise indicate that we have a conventional AFM, (ii) presence of at least one of $[C_2||C_{2}^{x}]$,  $[C_2||M_{x}]$, and $[C_2||C_{4}^{z}]$ spin group symmetries, ensuring altermagnetism, and (iii) absence of $T$, $PT$, $TC_{3,4,6}^{x,z}$, $TS_{4,6}^{x,z}$, $C_n^x$, $M_x$, $TM_z$, and $TC_2^z$ magnetic symmetries, which ensures non-zero $\sigma_{xy}$.  These conditions for obtaining 2D AMs with AHE are partially related. For example, $[C_2||P]$ and $PT$ occur simultaneously in collinear magnets. These conditions make observing AHE difficult in 2D AMs, and have largely remained unexplored.


%

The magnetic symmetries depend on the N\'eel vector orientation with respect to the crystal symmetry. The N\'eel vector serves as an order parameter for AFMs and offers a robust nonvolatile approach to modifying magnetic symmetries~\cite{jungwirth2016antiferromagnetic}. Since, AMs also have net zero magnetization, the N\'eel vector could serve as a natural way to modify AHE. For example, consider the case of a 2D AM with $[C_2||M_x]$ symmetry, where opposite spin sublattices are connected through a vertical mirror plane (see Fig.~\ref{p1}). When the N\'eel vector is perpendicular to the $x$-direction, the magnetic point group of the AM will contain $M_x$, leading to vanishing $\sigma_{xy}$ [Figs.~\ref{p1}(a) and ~\ref{p1}(b)]. However, aligning the N\'eel vector along the $x$-direction will lead to the breaking of the $M_x$ symmetry [Fig.~\ref{p1}(c)]. Hence, the presence of the N\'eel vector along the $x$-direction will allow $\sigma_{xy}$ [Fig.~\ref{p1}(d)]. It is worth mentioning that $M_x$ and $C_{2}^{x}$ impose the same condition on $\sigma_{xy}$ [see Table~\ref{t1}]. Therefore, the presence of [$C_2||C_{2}^{x}]$ will allow (forbid) $\sigma_{xy}$ when the N\'eel vector is parallel (perpendicular) to $x$ direction. Following the same approach, we have classified the 2D AMs based on whether the AHE is allowed for the different N\'eel vector orientations, spin-momentum coupling and their nontrivial spin layer group (SLG) (see Table~\ref{t2} and Sec. III of SM~\cite{SuMa}). Interestingly, $d$-wave  AMs with nontrivial SLG $^22/^2m$ and $i$-wave AMs with nontrivial SLG $^1\overline{3}{^2}m$ allows for an in-plane AHE, where $\boldsymbol{N}$, $\boldsymbol{J}_\textrm{H}$, and $\boldsymbol{E}$ lie in the $x-y$ plane. Additionally, AMs with nontrivial SLGs $^2m^2m^1m$, ${^2}4/{^1}m$, $^24/^1m^2m^1m$, $^14/^1m^2m^2m$, and $^16/^1m^2m^2m$ do not show an in-plane AHE due to the presence of $TC_2^z$ symmetry. In accordance to symmetry analysis, $\sigma_{xy}$ is always forbidden for 2D AMs when N\'eel vector points along the $z$-direction.
\begin{table*}[t]
	\caption{AHE in 2D AMs with different nontrivial spin-Layer group (SLG) symmetry. The superscripts $1$ and $2$ denote symmetry operations connecting atoms with the same and opposite spin magnetizations, respectively. Spin-momentum coupling is categorized based on the number of spin-degenerate nodal lines in the band dispersion. Specifically, $d$-, $g$-, and $i$-wave AMs exhibit 2, 4, and 6 spin-degenerate nodal lines passing through the $\Gamma$ point, respectively, in the ($k_x$-$k_y$) momentum space. The nontrivial spin group symmetry operations are highlighted for each case. Note that the spin group operations $[E||E]$ and $[E||P]$ are excluded as they impose no restrictions on the AHC. Symmetry-invariant terms describe the functional dependence of the AHC $\sigma_{xy}$ on the N\'eel vector components ($N_x, N_y, N_z$). The table also specifies whether $\sigma_{xy}$ is allowed when the N\'eel vector is perpendicular or parallel to the out-of-plane direction (z). For forbidden cases, the relevant magnetic symmetry responsible is also indicated. Symmetry invariants up to cubic in $N_i$ are included. Note that symmetry invariants may encompass higher-order terms, though their contributions are expected to be weak. Additionally, examples of well-known 2D AMs belonging to different SLGs are also tabulated. Entries marked with $[\star]$ denote examples from this work.}
	
	\centering
	\begin{tabular*}{\textwidth}{@{\extracolsep{\fill}}l c c c c c c }
		\hline
		\hline 
		\multirow{2}{*}{\centering \begin{tabular}{c}
				Nontrivial \\ 
				SLG
		\end{tabular}} & \multirow{2}{*}{\centering \begin{tabular}{c}
				Spin-momentum \\ 
				coupling
		\end{tabular}} & Nontrivial  &\multirow{2}{*}{\centering \begin{tabular}{c}
				Symmetry-allowed  \\ 
				$\sigma_{xy}$ terms
		\end{tabular}} & \multicolumn{2}{c}{$\sigma_{xy}$} &   \multirow{2}{*}{\centering \begin{tabular}{c} Examples \end{tabular}} \\
		\cline{5-6}
		& & spin group operations & & $N \perp z $  & $N || z$ &  \\
		\hline 
		$^22/^2m$& $d$-wave & [$C_2||C_{2}^{x}]$ & $N_x^{1,3}, N_xN_y^aN_z^b$ ($a$+$b$=$2$) & \ding{51} & \ding{55} ($C_{2}^{x}$) & RuF$_4$\cite{milivojevic2024interplay}, $d$-MnPSe$_3$ [$\star$]\\
		$^2m^2m^1m$& $d$-wave & [$C_2||C_{2}^{x}], [E||C_{2}^{z}]$ & $N_xN_yN_z$ & \ding{55} ($TC_2^z$) & \ding{55} ($C_{2}^{x}$) & MnTeMoO$_6$~\cite{zeng2024description} \\
		$^24/^1m$& $d$-wave & $[C_2||C_{4}^{z}], [E||C_{2}^{z}]$ & $N_x^2N_z-N_y^2N_z, N_xN_yN_z$  & \ding{55} ($TC_2^z$) & \ding{55} ($C_{2}^{x}$) & \\
		$^24/^1m^2m^1m$& $d$-wave & $[C_2||C_{4}^{z}], [E||C_{2}^{z},C_{2}^{x}]$ & $N_x^2N_z-N_y^2N_z$ & \ding{55} ($TC_2^z$) & \ding{55} ($C_{2}^{x}$) & VSe$_2$O~\cite{ma2021multifunctional}, CrO\cite{chen2023giant} \\
		$^14/^1m^2m^2m$& $g$-wave & $ [C_2||C_{2}^{x}], [E||C_{4}^{z}, C_{2}^{z}]$ & \textemdash &\ding{55} ($TC_2^z$) & \ding{55} ($C_{2}^{x}$) & VP$_2$H$_8$(NO$_4$)$_2$~\cite{zeng2024description}\\
		$^1\overline{3}{^2}m$& $i$-wave & $[C_2||C_{2}^{x}], [E||C_{3}^{z}]$ & $N_x^3-3N_xN_y^2$  & \ding{51}  & \ding{55} ($C_{2}^{x}$) & $i$-MnPSe$_3$ [$\star$] \\
		$^16/^1m^2m^2m$& $i$-wave & $[C_2||C_{2}^{x}], [E||C_{3}^{z}, C_{2}^{z}]$ & \textemdash & \ding{55} ($TC_2^z$)  & \ding{55} ($C_{2}^{x}$) & \\
		\hline \hline
	\end{tabular*}
	\label{t2}
\end{table*}

\begin{figure}[t]
	\includegraphics[width=8.6cm]{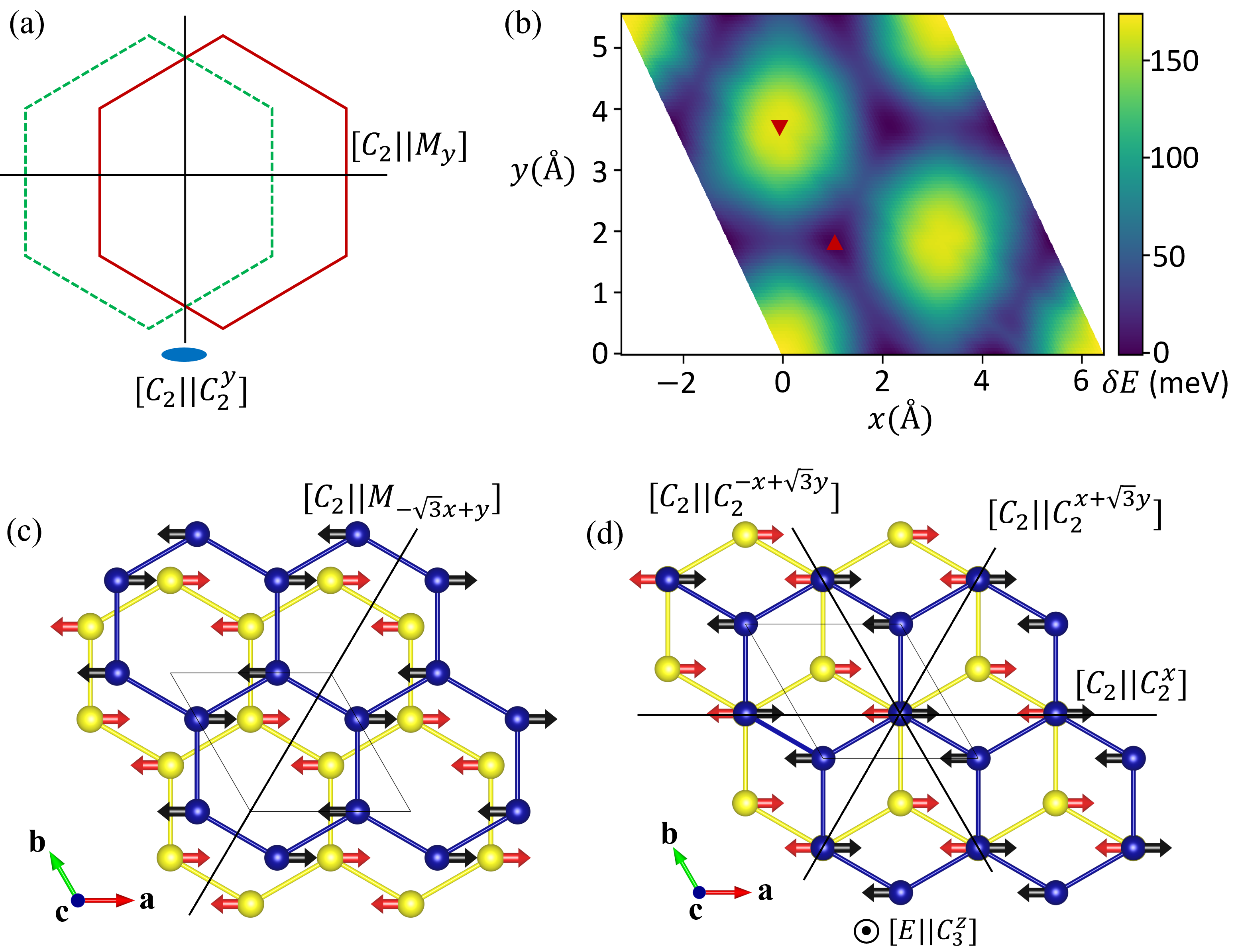}
	\caption{Generation of AM through bilayer stackings. (a) The upper layer is obtained by taking the horizontal mirror reflection of the lower layer followed by in-plane translation. The bilayer stacking may have $[C_2||M_y]$ and $[C_2||C_{2}^{y}]$-like spin group symmetries depending on the shifting vector and the constituent monolayer. (b) The energy distribution of the bilayer MnPSe$_3$ as a function of the shifting of the upper layer. (c) and (d) highlights the two high-symmetry stackings obtained by shifting the upper layer by $\frac{1}{3}\boldsymbol{a}$+$\frac{1}{3}\boldsymbol{b}$ ($d$-MnPSe$_3$) and $\frac{1}{3}\boldsymbol{a}$+$\frac{2}{3}\boldsymbol{b}$ ($i$-MnPSe$_3$), respectively, with representative N\'eel vector along $x$-direction. The blue and yellow spheres denote the Mn atoms from the top and bottom layers, respectively. We have omitted the P and Se atoms for clear illustration (see Fig. S2 in the SM~\cite{SuMa} for complete structure). The position of structures in (c) and (d) are also highlighted in potential energy surfaces in (b) with up- and down-triangle, respectively. The spin group symmetry operations are also indicated for each case.}
	\label{p2}
\end{figure}

\subsection{Bilayer AMs as prototypical candidates} We exemplify the symmetry predictions of AHE in the 2D AMs using DFT simulations. For this, we chose the prototypical experimentally-synthesized AFM materials --  MnPSe$_3$~\cite{ni2021imaging, li2014half, sivadas2015magnetic} and MnSe ~\cite{aapro2021synthesis, liu2023tunable, wang2023intrinsic, sheoran2024multiple} monolayers.  In what follows, we provide detailed results for MnPSe$_3$, while details of our calculations for MnSe are included in the supplemental material~\cite{SuMa}. Both  MnPSe$_3$ and MnSe monolayers form a large class of 2D materials with G and A type AFM order, respectively, and a N\'eel transition temperature of $\sim$75 K~\cite{yi2023exploring, aapro2021synthesis}. The $PT$ symmetry in MnPSe$_3$ and MnSe monolayers enforces spin-degeneracy and forbids the AHE effect (see Sec. IV of SM~\cite{SuMa}). To obtain AM, we break $PT$ symmetry of the monolayer MnPSe$_3$ and MnSe by using the bilayer stacking approach [see Fig~\ref{p2}(a)]~\cite{pan2024general}. We created different bilayer stacking by first taking the top layer to be the mirror reflection of the lower layer, followed by the translation of the upper layer in the basal plane. We have also used two different magnetic configurations with intralayer AFM, namely, $M$$\uparrow \downarrow \uparrow \downarrow$ and $M$$\uparrow \downarrow \downarrow \uparrow$ (the four arrows denote the magnetization directions of Mn atoms, with first two arrows being used for the lower layer and the last two arrows for the upper layer).  The configurations $M$$\uparrow \downarrow \uparrow \downarrow$ and $M$$\uparrow \downarrow \downarrow \uparrow$ are almost degenerate (differing by around 0.22\,meV), signifying the weak interlayer exchange interaction. Further, DFT calculations were performed for potential energy surfaces of various high symmetry stackings [see Fig~\ref{p2}(b) for MnPSe$_3$]. There are 6 degenerate lowest-energy configurations. These stackings are equivalent by symmetry and are obtained through translation of the upper layer by $\frac{1}{3}\boldsymbol{a}$, $\frac{1}{3}\boldsymbol{b}$, $\frac{2}{3}\boldsymbol{a}$, $\frac{2}{3}\boldsymbol{b}$, $\frac{1}{3}\boldsymbol{a}+\frac{1}{3}\boldsymbol{b}$, and $\frac{2}{3}\boldsymbol{a}+\frac{2}{3}\boldsymbol{b}$. Similarly, three high-energy stacking configurations, obtained through translation of the upper layer by $\boldsymbol{0}$, $\frac{1}{3}\boldsymbol{a}+\frac{2}{3}\boldsymbol{b}$, and $\frac{2}{3}\boldsymbol{a}+\frac{1}{3}\boldsymbol{b}$, are degenerate. Therefore, we take representative cases of bilayer MnPSe$_3$ obtained through $\frac{1}{3}\boldsymbol{a}+\frac{1}{3}\boldsymbol{b}$ [see Fig~\ref{p2}(c)] and $\frac{1}{3}\boldsymbol{a}+\frac{2}{3}\boldsymbol{b}$ [see Fig~\ref{p2}(d)] and we name those stackings as $d$-MnPSe$_3$ and $i$-MnPSe$_3$, respectively. The rationale behind this unconventional nomenclature is elaborated in the following paragraph. The most stable N\'eel ordering is $M$$\uparrow \downarrow \uparrow \downarrow$ and $M$$\uparrow \downarrow \downarrow \uparrow$ for $d$-MnPSe$_3$ and $i$-MnPSe$_3$, respectively, and therefore used in our simulations. Although, $i$-MnPSe$_3$ is not the most stable bilayer stacking [see Fig~\ref{p2}(d)], it serves a qualitative analysis of AHE in the materials sharing structural similarity. For instance, twisted bilayers (tb) of hexagonal materials (such as tb-NiCl$_2$~\cite{he2023nonrelativistic}, tb-MnBi$_2$Te$_4$~\cite{liu2024twisted}, and tb-MnPSe$_3$~\cite{sheoran2024nonrelativistic},) have the same nonmagnetic and magnetic point group as of $i$-MnPSe$_3$. However,  performing DFT+U+SOC simulations on twisted bilayers are computationally formidable task owing to supercell size.

\begin{figure}[ht]
	\includegraphics[width=8.6cm]{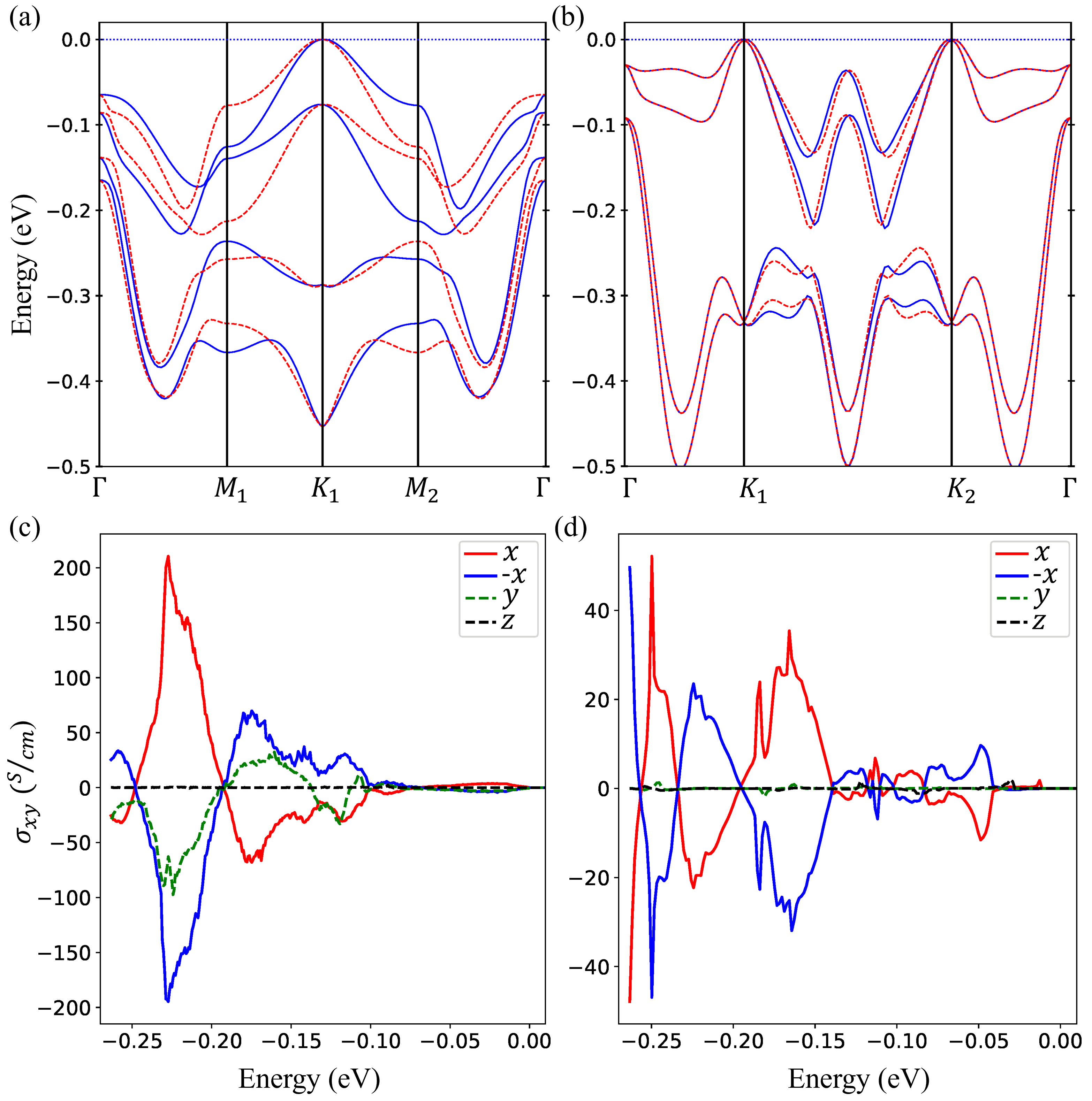}
	\caption{Valence bands of (a) $d$-MnPSe$_3$ and (b) $i$-MnPSe$_3$ bilayer without inclusion of SOC. The red dashes and blue dots represent spin-up and spin-down bands, respectively. Anomalous Hall conductivity $\sigma_{xy}$ of (c) $d$-MnPSe$_3$ and (d) $i$-MnPSe$_3$ as a function of Fermi energy for N\'eel vector pointing along different directions. The zero Fermi energy corresponds to valence band maximum. }
	\label{p3}
\end{figure}

We computed the electronic bands of $d$-MnPSe$_3$ and $i$-MnPSe$_3$ without SOC [see Figs.~\ref{p3}(a) and \ref{p3}(b)]. The spin degeneracy in the energy bands can be explained through spin group symmetry operations~\cite{vsmejkal2022beyond}. The presence of  [$C_2||O]$ symmetry leads to [$C_2||O]E(s,\boldsymbol{k}) = E(-s, O^{-1}\boldsymbol{k})$. Therefore, the bands are spin degenerate along the paths for which $O\boldsymbol{k}=\boldsymbol{k}$ or $O\boldsymbol{k}=\boldsymbol{k}+\boldsymbol{G}$, where $\boldsymbol{G}$ is a reciprocal lattice vector. For $d$-MnPSe$_3$, the spin-up and spin-down states are degenerate for the bands along the directions perpendicular and parallel to the mirror plane due to [$C_2||M_{-\sqrt{3}x+y}$] and [$\overline{C}_2||T$][$C_2||M_{-\sqrt{3}x+y}$], respectively (see Sec. V of SM~\cite{SuMa} for constant energy contours). AMs with two spin degenerate nodal lines are classified as $d$-wave AMs, and $d$-MnPSe$_3$ belong to that class~\cite{vsmejkal2022beyond}. Similarly, the energy bands in $i$-MnPSe$_3$ are spin degenerate along all possible high symmetry directions $\Gamma-M$, $\Gamma-K$, and $M-K$ due to presence of three [$C_2||C_2^x$]-type spin group symmetries [Fig.~\ref{p2}(d)] and their combinations with spin-only symmetry [$\overline{C}_2||T$]. This results in the $i$-wave altermagnetism in the $i$-MnPSe$_3$. Note that although bands are nondegenerate at the general $k$-point, the sum of spin splittings throughout the BZ, $\sum_{BZ}\mathcal{E}(s,\boldsymbol{k})-\mathcal{E}(-s,\boldsymbol{k})$, is zero for each case [see Figs.~\ref{p3}(a) and \ref{p3}(b)]. 

\begin{figure}[ht]
	\includegraphics[width=8.6cm]{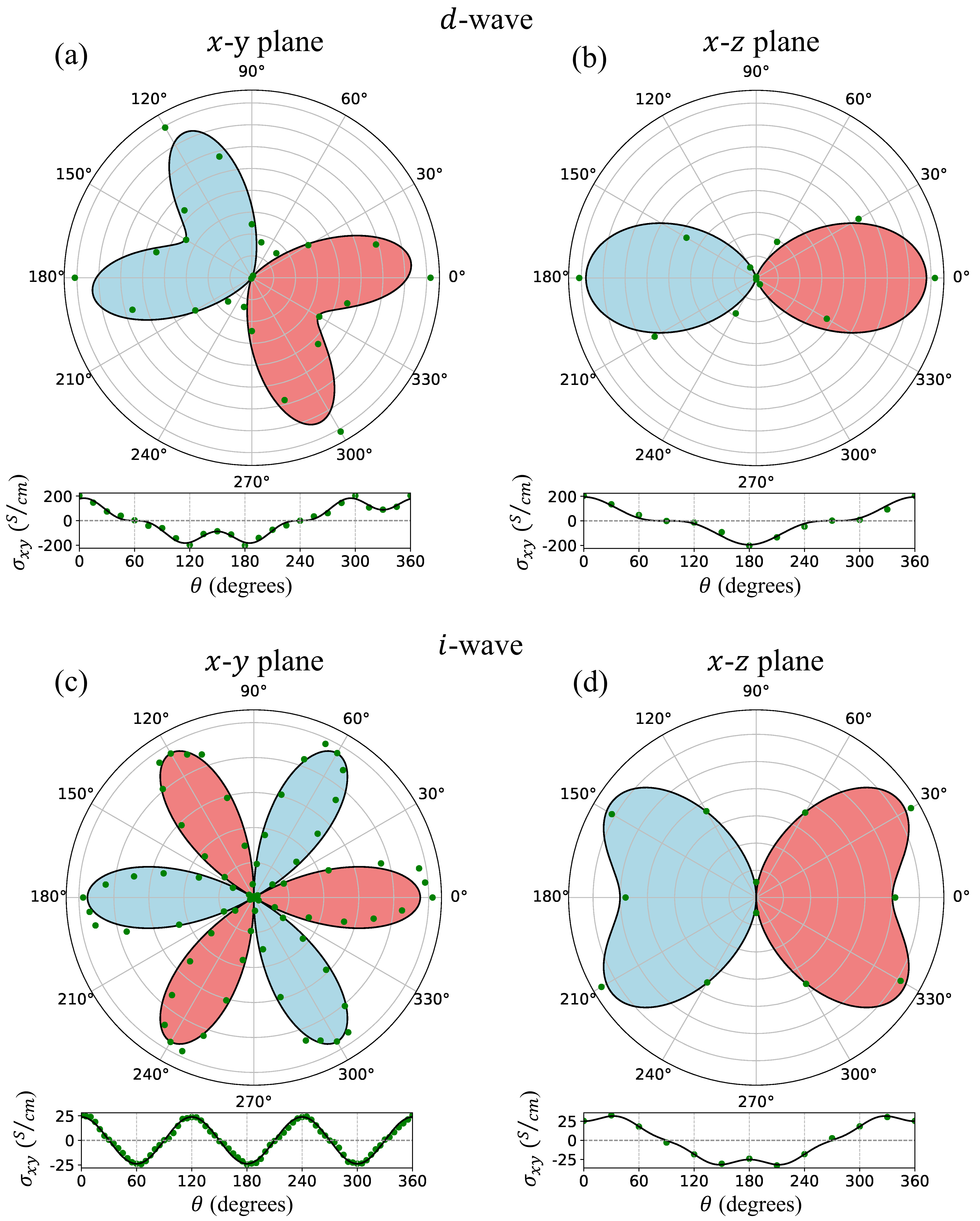}
	\caption{The anomalous Hall conductivity ($\sigma_{xy}$) of $d$-MnPSe$_3$ with $d$-wave AM as a function of the N\'eel vector orientation in (a) $x$-$y$ plane and (b) $x$-$z$ plane at $-$0.23 eV. Similarly, (c) and (d) show the AHC for $i$-MnPSe$_3$ with $i$-wave AM at $-$0.17 eV. The regions with positive and negative $\sigma_{xy}$ are highlighted with light red and light blue. The green dots represent the calculated DFT values while black curves are the fits using models in Eq.~\ref{e2}.}
	\label{p4}
\end{figure}

\subsection{The AHC and N\'eel vector relationship}  We calculate the AHC, $\sigma_{xy}$, as a function of the Fermi energy for $d$-MnPSe$_3$ and $i$-MnPSe$_3$ for different N\'eel vector, $\boldsymbol{N}$, orientations [Figs.~\ref{p3}(c) and \ref{p3}(d)]. The $\sigma_{xy}$ shows a strong dependence on the $\boldsymbol{N}$ orientation and is absent when $\boldsymbol{N}$ is along the $z$-direction. The anomalous Hall response is strongest for $\boldsymbol{N}$ along the $x$-direction. The $\sigma_{xy}$ for $d$-MnPSe$_3$ develops a strong peak of 202 S/cm around $-0.23$ eV, while $\sigma_{xy}$ of $i$-MnPSe$_3$ develops a slightly weaker peak of 25 S/cm around $-0.17$ eV. Note that we observe a very sharp peak for $i$-MnPSe$_3$ around $-0.25$ eV and it may be difficult to achieve experimentally as it requires precise control of the Fermi energy. Interestingly, $\sigma_{xy}$ is forbidden for $\boldsymbol{N}$ aligned along the $y$-direction for $i$-MnPSe$_3$, while it shows an intermediate response for $d$-MnPSe$_3$. To understand this dependence of $\sigma_{xy}$ on the N\'eel vector, we plot the anomalous Hall peaks for different $\boldsymbol{N}$ orientations in the $x$-$y$ and $x$-$z$ planes (see Fig.~\ref{p4}). For the $x$-$y$ plane, the AHE is absent in $d$-MnPSe$_3$ for the N\'eel vector along $60^\circ/240^\circ$ due to the presence of the vertical mirror $M_{-\sqrt{3}x+y}$ [Fig.~\ref{p4}(a)]. Similarly, the AHE is absent for $i$-MnPSe$_3$ when $\boldsymbol{N}$ is perpendicular to any of the two-fold rotation symmetries ($30^\circ, 90^\circ, 150^\circ, 210^\circ, 270^\circ$, and $330^\circ$)[Fig.~\ref{p4}(c)]. When $\boldsymbol{N}$ points in the $z$-direction, the AHE is forbidden due to presence of the $M_{-\sqrt{3}x+y}$ and $C_{2}^x$ in $d$-MnPSe$_3$ and $i$-MnPSe$_3$, respectively [see Figs.~\ref{p4}(b) and~\ref{p4}(d)]. 

Furthermore, we understand the periodicity by writing the general functional form of $\sigma_{xy}$ as $\sigma_{xy} (\boldsymbol{N}) = \sum_{m,n,r} \lambda_{mnr} N_x^{m}N_y^{n}N_z^{r}$, where $m$, $n$, and $r$ are whole numbers. Some components of $\lambda_{mnr}$ are forbidden due to symmetries. For instance, $m+n+r$ is never even as it would violate the Onsager reciprocity relation: $\sigma_{xy}(\boldsymbol{N}) = \sigma_{yx}(-\boldsymbol{N})$~\cite{landau2013electrodynamics}. We use the ``\textit{method of invariants}"~\cite{voon2009kp} [$O\sigma_{xy}(O \boldsymbol{N})= \sigma_{xy}({\boldsymbol{N}})$] to obtain symmetry allowed components $\lambda_{mnr}$. For this, we treat $\boldsymbol{N}$ as an extrinsic parameter instead of an intrinsic one~\cite{xiao2024anomalous, yuan2024nonrelativistic}. Recall that $\boldsymbol{N}$ transforms like $\boldsymbol{N}'=\pm \textrm{Det}(O) D(O) \boldsymbol{N}$, where $D(O)$ and $\textrm{Det}(O)$ are the matrix representation and determinant of $O$. The $+$ ($-$) sign is taken if $O$ exchanges the same (opposite) spin sublattices [see Figs~\ref{p1}(a) and~\ref{p1}(b)]. Table~\ref{t2} summarizes the symmetry allowed profiles of $\sigma_{xy}(\boldsymbol{N})$ for 2D AMs with different spin group symmetry operations and nontrivial SLG (see also Sec. III of SM~\cite{SuMa}). $d$-MnPSe$_3$ and $i$-MnPSe$_3$ have nontrivial SLG $^22/^2m$ and $^1\overline{3}{^2}m$, respectively, and the $\sigma_{xy}-\boldsymbol{N}$ relationship can be expressed as 
\begin{equation}
\begin{aligned}
	\sigma_{xy}^{^22/^2m} &= \lambda_{100}N_x+ \lambda_{300}N_x^3 +\lambda_{111}N_xN_yN_z 
	+ \lambda_{120}N_xN_y^2\\ &+ \lambda_{102}N_xN_z^2 \\
	\sigma_{xy}^{^1\overline{3}{^2}m} &= \lambda_{300}(N_x^3-3N_xN_y^2)+\lambda_{302}(N_x^3-3N_xN_y^2)N_z^2.
\end{aligned}
\label{e2}
\end{equation}
In Fig.~\ref{p4}, we fit the calculated DFT values of $\sigma_{xy}$ with the models in Eq.~\ref{e2}. Note that for $d$-MnPSe$_3$ the mirror plane is $M_{-\sqrt{3}x+y}$ and not $M_x$. Therefore, we have rotated the models accordingly while fitting. The models provide good agreement with the DFT results. However, the fit shows slight deviations from the DFT results, which may be  due to the structural distortions relative to the ideal symmetric configurations, and/or because we have ignored the higher-order terms in Eq.~\ref{e2}. Note that the $\sigma_{xy}$ of $d$-MnPSe$_3$ depends on linear and cubic terms in the N\'eel vector $\boldsymbol{N}$ [Fig~\ref{p4}(a)], while $i$-MnPSe$_3$ exhibits a purely cubic dependence on $\boldsymbol{N}$ [Fig~\ref{p4}(c)]. Consequently, the strength of AHC in $i$-MnPSe$_3$ is smaller than that in $d$-MnPSe$_3$, with three AHC-forbidden directions in the $x$-$y$ plane of the N\'eel vector space.

 The dependence of $\sigma_{xy}$ on $\boldsymbol{N}$ is derived for all 2D altermagnetic nontrivial SLGs in Table~\ref{t2}. The AHE is forbidden for all $g$-wave 2D AMs due to the presence of $[E||C_4^z]$ symmetry, while the $[E||C_2^z, C_3^z]$ symmetries suppress the AHE in $i$-wave 2D AMs with the nontrivial SLG $^16/^1m^2m^2m$. The presence of $[E||C_2^z]$, $[E||C_4^z]$, or $[C_2||C_4^z]$ leads to the presence of magnetic $TC_2^z$ symmetry, which forbids the in-plane AHE. Most commonly, 2D magnets have an easy axis of magnetization that lies completely in-plane~\cite{rybak2024magneto, sheoran2024multiple} or out-of-plane~\cite{li2019intrinsic,webster2018strain}, and in such a scenario, the AHE can be observed for only $^22/^2m$ and $^1\overline{3}{^2}m$ out of the seven nontrivial SLGs. Interestingly, the dependence of $\sigma_{xy}$ on the N\'eel vector space  is similar to that of the Berry curvature multipoles on $k$-space~\cite{zhang2023higher} and the spin magnetization multipoles on real space~\cite{bhowal2024ferroically}.
 
The quantum origin of the large AHC lies in the SOC-induced avoided crossings, which act as opposite poles of Berry curvature (see Sec. VI of SM~\cite{SuMa}). The highest $\sigma_{xy}$ is achieved if the Fermi energy lies in the middle of the poles. In the case of the antiferromagnetic MnSe bilayer, with the translation of one layer relative to the other by $\frac{1}{3}\boldsymbol{a}+\frac{1}{3}\boldsymbol{b}$, we obtain effects similar to those in the $d$-MnPSe$_3$ bilayer, including altermagnetism (see Sec. VII of SM~\cite{SuMa}). Note that bilayer stacking of AFMs may also lead to weak ferromagnetism~\cite{jo2024weak, roig2024quasi} where the opposite spin sublattices are not connected by any symmetry. For instance, the polar stacking of bilayer MnSe with basal plane translated by $\frac{1}{3}\boldsymbol{a}+\frac{2}{3}\boldsymbol{b}$ leads to weak ferromagnetism in the structure. In such cases, the anomalous Hall response is FM-like, where the AHE is also observed in the plane perpendicular to N\'eel vector. The weakly FM bilayer MnSe shows $\sigma_{xy}$ when the N\'eel vector points along the $z$-direction similar to conventional FMs~\cite{cao2023switchable, chong2024intrinsic} (see Sec. VIII of SM~\cite{SuMa}) and differ from the AHE in 2D AMs. 

\section{Summary} We have established the principles for obtaining the AHE in 2D AMs.  Our analysis shows that $TC_2^z$ and $C_2^x$ are the most common magnetic point group symmetries that suppress the AHE in 2D AMs. Although the AHE is a relativistic effect, spin group symmetries are effective in explaining the unconventional periodicity in the in-plane anomalous Hall response, which can be used to detect the altermagnetic order and N\'eel vector reversal. We also show that the AHE is forbidden for the $g$-wave 2D AMs. Our symmetry predictions are supported by first-principles DFT simulations for bilayer MnPSe$_3$ with two-different layer-stacking geometries used as prototypical candidates for the $d$- and $i$-wave AMs. Overall, 2D AMs are promising for the miniaturization of spintronic memory devices, with the N\'eel vector serving as the write-in mechanism and in-plane AHE for read-off of the information. For the purpose of application in spintronics, 2D AMs that have an in-plane easy axis of magnetization will form a suitable choice.  The spin-momentum coupling, and hence, the altermagnetic order can be controlled in twisted magnetic bilayers, depending on the symmetry of the constituent monolayers~\cite{liu2024twisted},  providing an exceptional platform to achieve the AHE in experiments.  Additionally, external parameters such as electric fields or strain may induce AHE in otherwise forbidden AMs~\cite{lejman2019magnetoelastic}. The approach provided here can be extended to other similar transverse effects, such as the anomalous Nernst effect~\cite{mizuguchi2019energy} and the nonlinear Hall effect~\cite{du2021nonlinear}. We anticipate that the theoretical findings of this work will enrich the field of altermagnetic spintronics~\cite{vsmejkal2022emerging}.

\section*{Acknowledgments}  
This material is based upon work supported by the Air Force Office of Scientific Research under award number FA9550-23-1-0679 and the National Science Foundation Grant No. DMR-1752840. This work used the Expanse cluster at SDSC through allocation PHY180014 from the Advanced Cyberinfrastructure Coordination Ecosystem: Services \& Support (ACCESS) program, which is supported by National Science Foundation Grants Nos. 2138259, 2138286, 2138307, 2137603, and 2138296, and Maryland Advanced Research Computing Center. 

%


\end{document}


\title{Supplemental Material for\\
	``Spontaneous Anomalous Hall Effect in Two-Dimensional Altermagnets"}
\author{Sajjan Sheoran}
\author{Pratibha Dev} 
\affiliation{Department of Physics and Astronomy, Howard University, Washington D.C., USA}
\pacs{}
\maketitle
\begin{enumerate}[\bf I.]
\item 2D \MakeLowercase{altermagnetism}: group theoretic considerations	
\item Symmetry constraints on AHE 
\item AHE and N\'{e}el vector relationship
\item Monolayer MnPSe$_3$ and MnSe
\item Altermagnetism in bilayer MnPSe$_3$
\item Origin of AHE in bilayer MnPSe$_3$
\item AHE in altermagnetic bilayer MnSe
\item AHE in weakly ferromagnetic bilayer MnSe

\end{enumerate}

\section{2D A\MakeLowercase{ltermagnetism: Group Theoretic Considerations}}

{The main text describes symmetry considerations that lead to the emergence of 2D altermagnetism.  In this subsection, we provide greater details of how group theoretic considerations were used to obtain our results.} 

If the Hamiltonian of a system remains invariant under a certain symmetry ($O$), it transforms the energy eigenstate ($E$) as $OE(s, \boldsymbol{k}) = E(s', \boldsymbol{k}')$, where $s' = Os$ and $\boldsymbol{k}' = O\boldsymbol{k}$. In the nonrelativistic case, spin and real space are decoupled.  In this case, the symmetry transformation, $O$, can be written as $O = [O_1||O_2]$, where $O_1$ and $O_2$ act only on the spin and real spaces, respectively. Generally, $[O_1||O_2]$ can take one of four forms: $[E||E]$, $[O_1||E]$, $[E||O_2]$, and $[O_1||O_2]$. Here, $[E||E]$ represents the trivial identity in both spaces, $[O_1||E]$ corresponds to spin-only symmetry, $[E||O_2]$ represents crystallographic symmetry (nonmagnetic point group), and $[O_1||O_2]$ represents nontrivial spin group symmetry. Collinear magnets can exhibit spin-only symmetries, such as $C_\infty$ (arbitrary rotation around the collinear spin axis) and $\overline{C}_2$ (a two-fold rotation around an axis perpendicular to the collinear spin axis, followed by inversion) in addition to the trivial identity. The symmetry $C_\infty$ allows the separation of spin-up and spin-down bands. Meanwhile, spin-space inversion in $\overline{C}_2$ occurs with the time-reversal operation, which also acts on the real space. As a result, collinear magnets always exhibit $[\overline{C}_2||T]$ symmetry. For collinear magnets without spin-orbit coupling (SOC), the general symmetry operation can be written as:
\[
([E||E] + [C_\infty||E] + [\overline{C}_2||T] + [C_\infty||E][\overline{C}_2||T])[O_1||O_2],
\]
where the terms inside the first set of brackets represent the spin-only group, and $[O_1||O_2]$ is a nontrivial spin group symmetry dependent on the crystal and magnetic structure. For example, if opposite spin sublattices are connected by inversion ($P$) or translation ($\tau$), the nontrivial spin group symmetry is $[C_2||P]$ or $[C_2||\tau]$, where $C_2$ represents a two-fold rotation about an axis perpendicular to the collinear spin axis. The $C_2$ acts on spin-space same as spin-space inversion [the difference being that it is not accompanied by time-reversal ($T$)]. The $[\overline{C}_2||T][C_2||P]$ and $[C_2||\tau]$ symmetries transform energy eigenstates as follows:
\begin{equation}
[\overline{C}_2||T][C_2||P]E(s, \boldsymbol{k}) =[\overline{C}_2||T]E(-s, -\boldsymbol{k})=E(-s, \boldsymbol{k})
\end{equation}
\begin{equation}
[C_2||\tau]E(s, \boldsymbol{k}) = E(-s, \boldsymbol{k})
\end{equation}
Therefore, $[C_2||P]$ and $[C_2||\tau]$ lead to spin degeneracy at arbitrary an $\boldsymbol{k}$-point. For the 2D materials, $\boldsymbol{k}=k_xi+k_yj$ ($k_z=0$). The transformations of $[C_2||M_z]$ and $[\overline{C}_2||T][C_2||C_{2z}]$ act on energy eigenstates as follows:
\begin{equation}
[C_2||M_z]E(s, \boldsymbol{k}) = E(-s, \boldsymbol{k})
\end{equation}
\begin{equation}
[\overline{C}_2||T][C_2||C_{2z}]E(s, \boldsymbol{k}) = [\overline{C}_2||T]E(-s, -\boldsymbol{k})=E(-s, \boldsymbol{k})
\end{equation}
Here, the operator $M_z$ represents mirror symmetry about the $ z = 0 $ plane, while $ C_{2z} $ denotes a two-fold rotation about the $ z $-axis in real space. Consequently, the symmetries $ [C_2 \Vert C_{2z}] $ and $ [C_2 \Vert M_z] $ enforce spin degeneracy in 2D materials. To break spin degeneracy in 2D materials, opposite spin sublattices must not be related by $P$, $ \tau $, $ M_z $, or $ C_{2z} $. 

For 2D altermagnetism to emerge, there must be specific symmetries that connect opposite spin sublattices. While, in principle, any symmetry except $ P $, $ \tau $, $ M_z $, and $ C_{2z} $ can connect opposite spin sublattices, not all such symmetries are practical in 2D materials. For example, 2D materials cannot exhibit $ C_{3}^x $ symmetry. Similarly, although $ C_3^z $ symmetry is allowed in 2D materials, it cannot connect opposite spin sublattices, as it would lead to
$
[C_2 \Vert C_3^z][C_2 \Vert C_3^z][C_2 \Vert C_3^z] = [C_2 \Vert E],
$ which means opposite spin sublattices are connected by identity (not possible for magnets).  Similarly it can be shown that the only symmetries that can connect opposite spin sublattices in 2D materials are $ C_2^x $, $ C_4^z $, and $ M_x $. In spin group notation, the corresponding symmetries are $ [C_2 \Vert C_2^x] $, $ [C_2 \Vert C_4^z] $, and $ [C_2 \Vert M_x] $. It is worth mentioning that symmetry $[C_2||S_4^{z}]$ may also connect opposite spin sublattices, however, it is redundant as it can be written as combination of $[E||P]$ and $[C_2||C_4^{z}]$ and we are concerned only about generators.

\section{S\MakeLowercase{ymmetry constraints on} AHE}
{In what follows, we provide our procedure for determining the constraints on the AHC tensor components, which are imposed by unitary and antiunitary symmetry operations (see Table I of the main text).}  The AHC is expressed as:  
\begin{equation}  
	\sigma_{\alpha\beta} = -\epsilon_{\alpha\beta\gamma} \frac{e^2}{\hbar} \int \frac{d \boldsymbol{k}}{(2\pi)^3} \sum_n f(\mathcal{E}_{n, \boldsymbol{k}}) \Omega_{\gamma}(n, \boldsymbol{k}),  
	\label{E1}  
\end{equation}  
where \(\Omega_{\gamma}(n, \boldsymbol{k})\) is the Berry curvature of band \(n\) at the \(\boldsymbol{k}\)-point, $\mathcal{E}_{n, \boldsymbol{k}}$ is the energy eigenvalue, $\epsilon_{\alpha \beta \gamma}$ is the Levi-Civita symbol, and $f(\mathcal{E}_{n, \boldsymbol{k}})$ is the Fermi-Dirac distribution function. The Berry curvature and \(\boldsymbol{\sigma}_H\) (\(=\sigma_{zy}, \sigma_{xz}, \sigma_{yx}\)) transform as pseudovectors. Also, the AHC tensor is antisymmetric in nature, \(\sigma_{ij} = -\sigma_{ji}\). For example, under \(T\) symmetry, \(T\sigma_{xy}(\boldsymbol{k}) = -\sigma_{xy}(-\boldsymbol{k})\). Therefore, the summation in Eq.~\ref{E1} evaluates to zero, even though the Berry curvature is not zero at every \(\boldsymbol{k}\)-point. The \(T\) symmetry is satisfied for nonmagnetic materials; a well-known example is MoS\(_2\). Similarly, \(PT\) symmetry leads to  
\[
PT\boldsymbol{\Omega}(\boldsymbol{k}) = -P\boldsymbol{\Omega}(-\boldsymbol{k}) = -\boldsymbol{\Omega}(\boldsymbol{k}),
\]  
which results in zero Berry curvature at every \(\boldsymbol{k}\)-point [$\boldsymbol{\Omega}(\boldsymbol{k})=0$], thereby forbidding the anomalous Hall effect (AHE).  

Let us consider the case of \(M_x\) symmetry. The \(M_x\) symmetry transforms the Berry curvature as  
\[
M_x\Omega(\boldsymbol{k}) = (\Omega_x, -\Omega_y, -\Omega_z)(-k_x, k_y, k_z).
\]  
Consequently, \(\sigma_{zy}\) is the only nonzero component in this case. Alternatively, \(M_x\) transforms \( \boldsymbol{\sigma}_{H}\) as   \(M_x (\sigma_{zy}, \sigma_{xz}, \sigma_{yx})= (\sigma_{zy}, -\sigma_{xz}, -\sigma_{yx})\), which implies that only \(\sigma_{yz} \neq 0\). {Using a similar procedure for the rest of the symmetry operations, one can build Table I of the main text. }

\section{AHE \MakeLowercase{and} N\MakeLowercase{\'{e}el vector relationship}}
Figure 4 of the main text shows the functional dependence of the AHE on the N\'{e}el vector. Writing the general form of the AHC as $\sigma_{xy}=\sigma_{xy}(\boldsymbol{N})$, one can obtain the functional dependence of the AHE on the N\'{e}el vector as: 
\begin{equation}
	\sigma_{xy} (\boldsymbol{N}) = \sum_{m,n,r} \lambda_{mnr} N_x^{m}N_y^{n}N_z^{r},
\end{equation}
where, $\lambda_{mnr}$ are the coefficients that determine the contribution of each term. Some terms are forbidden by symmetry. For example, $\sigma_{xy}$ cannot be even in $\boldsymbol{N}$ as it would violate the {Onsager reciprocity relation:} $\sigma_{xy}(\boldsymbol{N})=-\sigma_{xy}(-\boldsymbol{N})$. The N\'{e}el vector is determined by the magnetization of sublattices. {Therefore, the N\'{e}el vector transforms as a pseudovector under the space group operation, $O$. However, it acquires an additional sign ($\pm$), depending on whether transformation exchanges opposite ($-$) or same ($+$) spin sublattices (see Fig. 1 of main text). Mathematically, this can be expressed as:}
\begin{equation}
	O\boldsymbol{N}=\pm Det(O)D(O)\boldsymbol{N}.
\end{equation}
{where $Det(O)$ is the determinant of the matrix [$D(O)$] for the symmetry operation $O$.} For instance, if the opposite spin sublattices are connect by Inversion ($P$), then $P\boldsymbol{N}=(-1)(-1)(-I_{3\times3})\boldsymbol{N}$ and therefore $P\boldsymbol{N}=-\boldsymbol{N}$. Also, $P$ transforms $\sigma_{H}$ as $P\boldsymbol{\sigma}_{H}=(-1)(-I_{3\times3})\boldsymbol{\sigma}_{H}$, which mean $P\sigma_{xy}=\sigma_{xy}$. Therefore, when the opposite spin sublattices are connect by inversion, all of the $\lambda_{mnr}$ coefficients are symmetry forbidden, suppressing the AHE. This can also be understood from the fact that the system has $PT$ ($[C_2||P]$ in spin group notation) symmetry and so {linear} AHE is not allowed (although, higher-order AHE may be allowed). Similar arguments can be used to show that if inversion symmetry connects the same spin sublattices ($[E||P]$), it does not impose any condition on the Hall effect. These observation are consistent with the Table I of the main text.
 
For the case of 2D altermagnetism (see Sec I of SM), the allowed spin group symmetries are $[C_2 || C_2^x, M_x, C_4^z]$ (which connect opposite spin sublattices) and $[E || E, P, C_2^z, M_z, C_2^x, M_x, C_3^z, C_4^z]$ (which connect same-spin sublattices). Note that the in-plane direction is chosen along the $x$-axis. Other symmetry operations, such as $[C_2||{S}_4^z$], $[E||{S}_4^z]$, and $[E||{S}_3^z]$ can be generated from the aforementioned operations, and thus are safely ignored.

Out of 80 layer groups only a few can have spin group symmetry $[C_2 || C_2^x, M_x, C_4^z]$. The possible layer groups are $2/m$, $mmm$, $4/m$, $4/mmm$, $3m$, and $6/mmm$. Note that layer groups containing screw axes and glide planes are ignored, although they can also be included. For example, observations for $2/m$ should also apply to $2_1/m$, $2/b$, and $2_1/b$. For a particular layer group, symmetry operations will either connect opposite spin sublattices or same-spin sublattices. Operations connecting same- and opposite-spin sublattices are highlighted with superscripts, $1$ and $2$, respectively, which are added before the symmetry operation. For example, in $^2 4/^1 m^2 m^2 m$, the fourfold rotation connects opposite spin sublattices, the horizontal mirror connects same-spin sublattices, and the two vertical mirror planes connect opposite spin sublattices. There can be seven nontrivial spin layer groups, namely $^2 2/^2 m$, $^2 m^2 m^1 m$, $^2 4/^1 m$, $^2 4/^1 m^2 m^1 m$, $^2 4/^1 m^2 m^2 m$, $^1 \overline{3}{^2}m$, and $^1 6/^1 m^2 m^2 m$. Therefore, we derive the relationship between $\sigma_{xy}$ and $\boldsymbol{N}$ for these seven nontrivial spin layer groups in Table II of the main text. Here, we highlight the case of $^2 4/^1 m^2 m^1 m$ (see Table S1). The generators of $^2 4/^1 m^2 m^1 m$ are $[E || E]$, $[E || P]$, $[E || C_2^z]$, $[E || C_2^x]$, and $[C_2 || C_4^z]$. The invariant common to all symmetry operations is $N_x^2 N_z - N_y^2 N_z$. Therefore, for the $^2 4/^1 m^2 m^1 m$ nontrivial spin layer group, we obtain
\[
\sigma_{xy}^{^2 4/^1 m^2 m^1 m} = \lambda_{201} (N_x^2 N_z - N_y^2 N_z).
\]

\begin{table}[h!]
	\centering
	\caption{Summary of invariants under the symmetry operations of $^2 4/^1 m^2 m^1 m$.}
	\begin{tabular}{|c|c|l|}
		\hline \hline
		\textbf{Symmetry Operation} & \textbf{Constraint on $\sigma_{xy}(N_x, N_y, N_z)$} & \textbf{Symmetry Invariants} \\ \hline
		$[E || E]$ & $\sigma_{xy}(N_x, N_y, N_z) = \sigma_{xy}(N_x, N_y, N_z)$ & $N_x^{m}N_y^{n}N_z^{r} (m+n+r=\textrm{odd})$ \\ \hline
    	$[E || P]$ & $\sigma_{xy}(N_x, N_y, N_z) = \sigma_{xy}(N_x, N_y, N_z)$ & $N_x^{m}N_y^{n}N_z^{r} (m+n+r=\textrm{odd})$ \\ \hline
		$[C_2 || C_4^z]$ & $\sigma_{xy}(N_x, N_y, N_z) = \sigma_{xy}(-N_y, N_x, -N_z)$ & $N_x^2 N_z - N_y^2 N_z$ \\ \hline
		$[E || C_2^z]$ & $\sigma_{xy}(N_x, N_y, N_z) = \sigma_{xy}(-N_x, -N_y, N_z)$ & $N_z$, $N_x^2 N_z$, $N_y^2 N_z$ \\ \hline
		$[E || C_2^x]$ & $\sigma_{xy}(N_x, N_y, N_z) = -\sigma_{xy}(N_x, -N_y, -N_z) = \sigma_{xy}(-N_x, N_y, N_z)$ & $N_y$, $N_z$, $N_x^2 N_y$, $N_x^2 N_z$, $N_y^2 N_z$, $N_y N_z^2$ \\ \hline \hline
	\end{tabular}
	\label{tab:invariants}
\end{table}

\section{M\MakeLowercase{onolayer} M\MakeLowercase{n}PS\MakeLowercase{e}$_3$ \MakeLowercase{and} M\MakeLowercase{n}S\MakeLowercase{e}}
\begin{figure}[H]
	\centering
	\includegraphics[width=14cm]{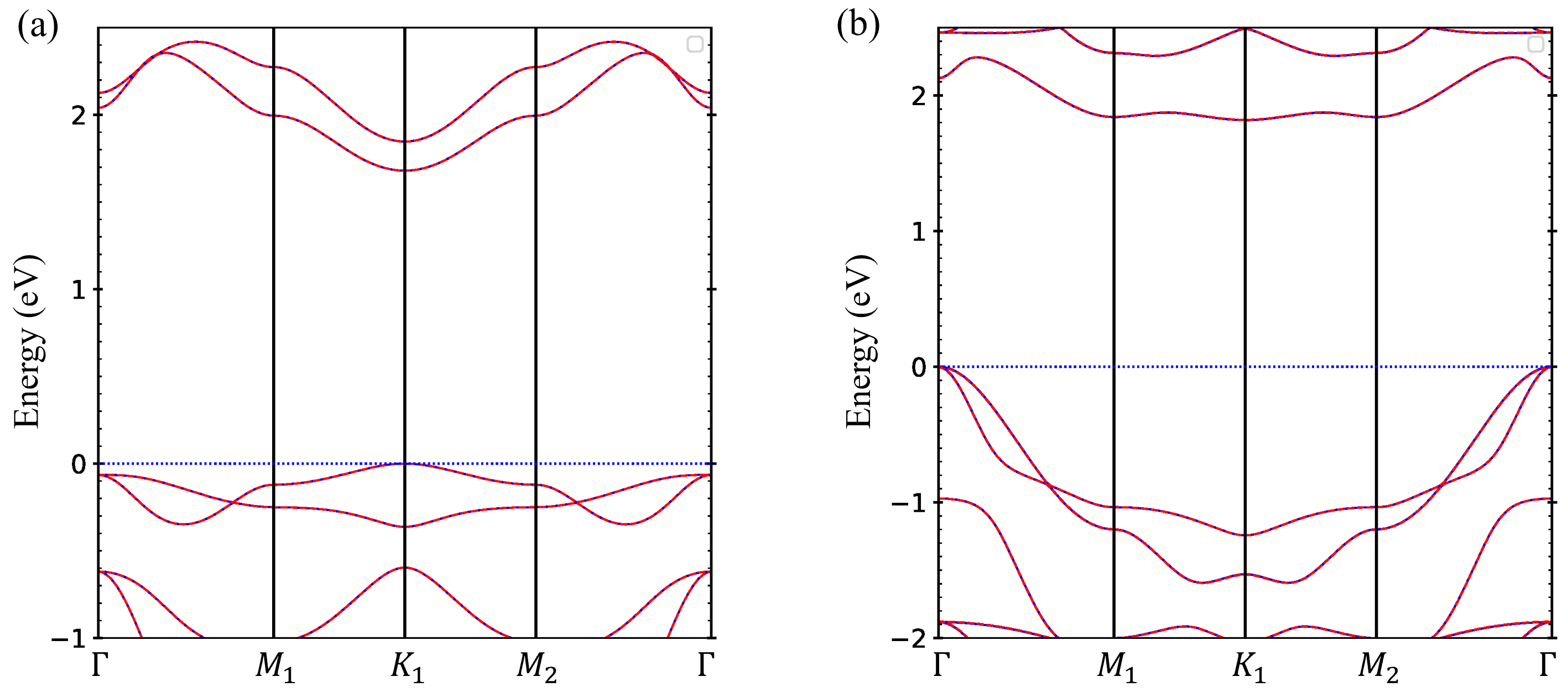}
	\caption{Spin-polarized band structure of (a) monolayer MnPSe$_3$ and (b) monolayer MnSe without SOC. The red and blue curves denote the spin-up and spin-down bands, respectively. The bands are degenerate throughout the BZ due to spin group symmetry $[C_2||P]$. Similarly, the bands remain degenerate throughout the BZ with the inclusion of SOC due to $PT$ symmetry (not shown here; see, for example, Refs.~\cite{sivadas2016gate, liu2023tunable}). We make use of PyProcar~\cite{herath2020pyprocar} to plot the band structures.
	}
	\label{p1}
\end{figure}

The opposite spin sublattices in monolayer MnPSe$_3$ and MnSe are connected by $[C_2||P]$ symmetry. Therefore, the bands are degenerate throughout the BZ (see Fig.~\ref{p1}). Similarly, the bands remain degenerate throughout the BZ with the inclusion of SOC due to $PT$ symmetry (see Sec.~I). $PT$ symmetry also suppresses the AHE in both cases (see Sec.~II). Note that we have chosen these two materials as they exhibit tilted Dirac cones near the valence band maximum along the $\Gamma-M$ path. These tilted Dirac cones may act as a source of strong Berry curvature when $PT$ symmetry is broken (see, for example, Ref.~\cite{wang2021intrinsic}).

\begin{figure}[H]
	\centering
	\includegraphics[width=13cm]{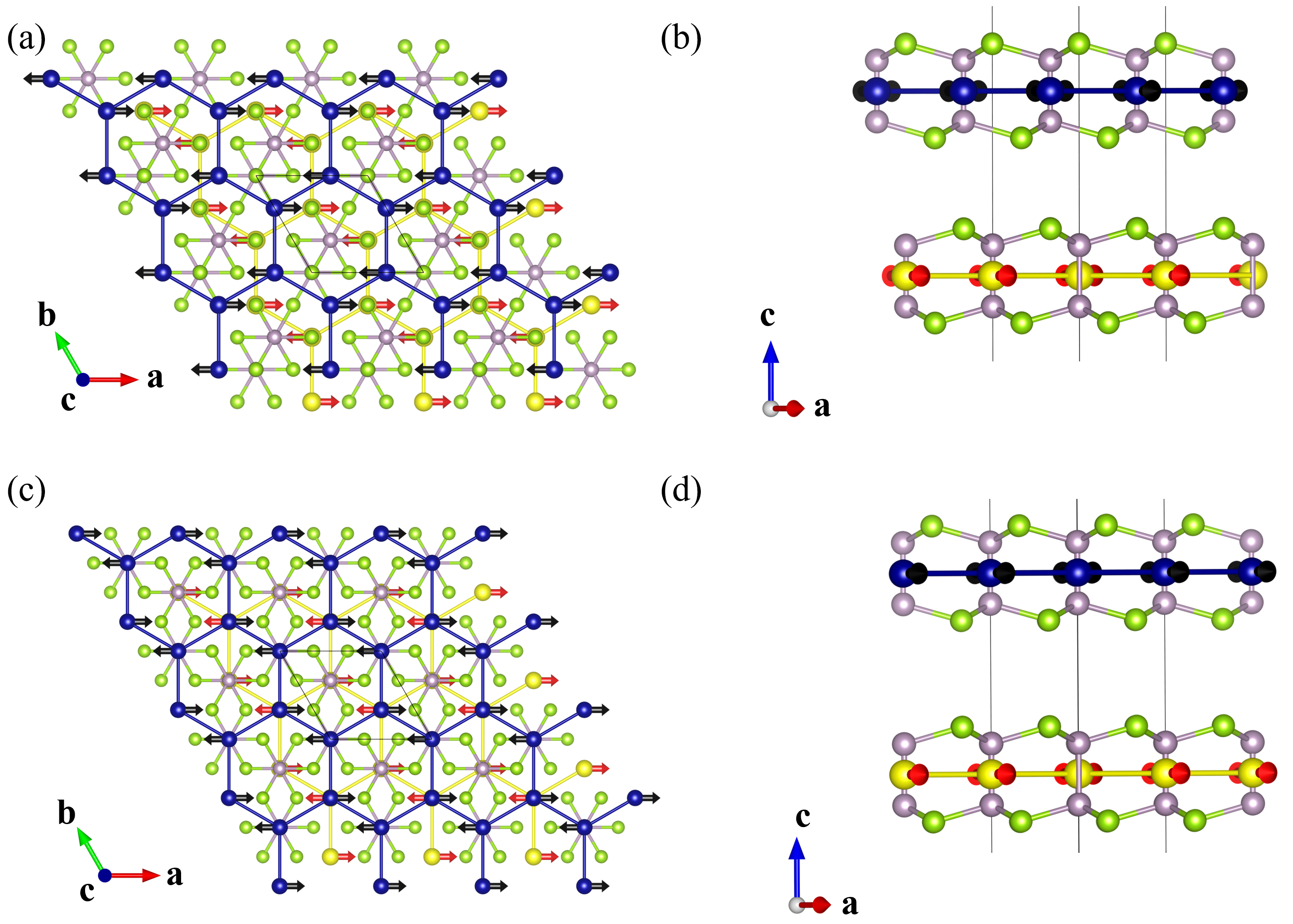}
	\caption{The top and side views of bilayer $d$-MnPSe$_3$ [(a) and (b)] and $i$-MnPSe$_3$ [(c) and (d)]. The blue and yellow spheres denote the Mn atoms from the top and bottom layers, respectively. The green ad grey spheres represent P and Se atoms, respectively.
	}
	\label{p2}
\end{figure}

\section{A\MakeLowercase{ltermagnetism in bilayer} M\MakeLowercase{n}PS\MakeLowercase{e}$_3$}
\begin{figure}[H]
	\centering
	\includegraphics[width=16cm]{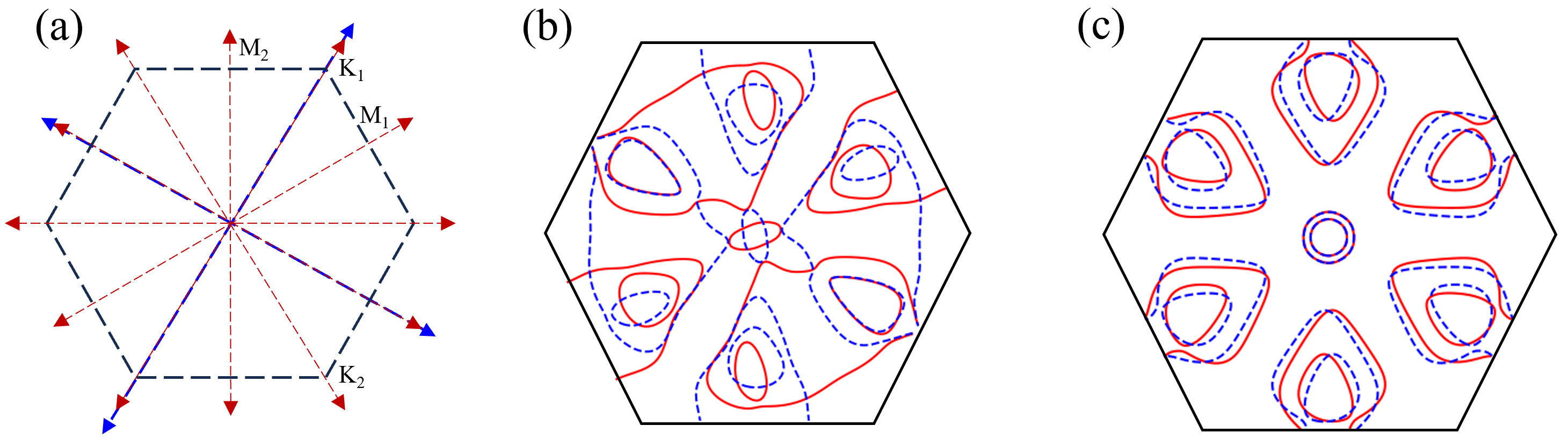}
	\caption{
			(a) The Brillouin Zone of the hexagonal unit cell, showing key symmetry points used in the band structure plots. (b) Constant energy contour at $-0.23$ eV for $d$-MnPSe$_3$ and (c) constant energy contour at $-0.17$ eV for $i$-MnPSe$_3$. Red solid lines and blue dashed lines represent spin-up and spin-down bands, respectively. Spin-degenerate nodal lines are highlighted in (a), with blue and red arrows marking the spin-degenerate lines for $d$-MnPSe$_3$ and $i$-MnPSe$_3$, respectively.}
	\label{p3}
\end{figure}

Figure ~\ref{p2} shows the two possible high symmetry stackings of bilayer MnPSe$_3$, named $d$-MnPSe$_3$ and $i$-MnPSe$_3$. To understand the altermagnetism in bilayer MnPSe$_3$, we plot constant energy surfaces for $d$-MnPSe$_3$ and $i$-MnPSe$_3$ in Fig.~\ref{p3}. The $d$-MnPSe$_3$ and $i$-MnPSe$_3$ exhibit 2 and 6 spin-degenerate nodal lines passing through the $\Gamma$ point, respectively. For $d$-MnPSe$_3$, this can be understood from the fact that $d$-MnPSe$_3$ has $[C_2||M_{-\sqrt{3}x+y}]$ symmetry, which enforces spin degeneracy along the $\Gamma$-$K_1$ path. The $[\overline{C}_2||T][C_2||M_{-\sqrt{3}x+y}]$ symmetry enforces spin degeneracy along the other spin-degenerate line. In a similar way, six spin group symmetries for $i$-MnPSe$_3$—three being $[C_2||C_2^x]$, $[C_2||C_2^{x+\sqrt{3}y}]$, and $[C_2||C_2^{-x+\sqrt{3}y}]$, and three being their combinations with $[\overline{C}_2||T]$—enforce spin degeneracy along the $\Gamma$-$K$ and $\Gamma$-$M$ directions (six in total).  The nonrelativistic dispersions in $d$-MnPSe$_3$ and $i$-MnPSe$_3$ are classified as $d$-wave and $i$-wave altermagnetism, respectively. Similarly, $g$-wave altermagnetism exhibits 4 spin-degenerate nodal lines crossing the $\Gamma$ point in the 2D Brillouin zone~\cite{vsmejkal2022beyond}.

\section{O\MakeLowercase{rigin of} AHE  \MakeLowercase{in bilayer} M\MakeLowercase{n}PS\MakeLowercase{e}$_3$}
\begin{figure}[H]
	\centering
	\includegraphics[width=17cm]{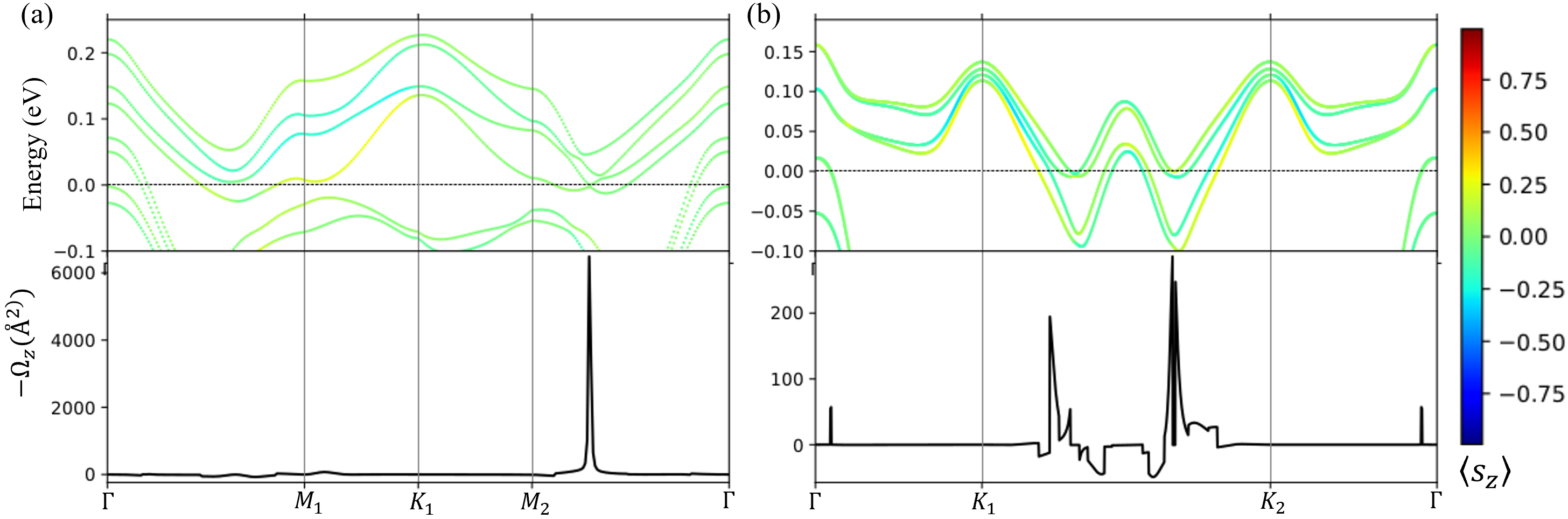}
	\caption{The spin-polarized band structure (upper panel) and total Berry curvature $-\Omega_z(\boldsymbol{k})$ (lower panel) for (a) $d$-MnPSe$_3$ and (b) $i$-MnPSe$_3$ with SOC and magnetization along the $x$-direction. The spin-polarized bands are colored according to the spin expectation along the $z$-direction. The Fermi energy is shifted to the energy level at which significant anomalous Hall conductivity is observed, as discussed in the main text (-0.23 eV and -0.17 eV below the valence band maximum for $d$-MnPSe$_3$ and $i$-MnPSe$_3$, respectively).}
	\label{p4}
\end{figure}

The Berry curvature shows peaks at the band anticrossings in FMs~\cite{singh2023kagome}, AFMs~\cite{chen2014anomalous}, and AMs~\cite{vsmejkal2020crystal}. It is not surprising then that the contribution to AHC in $d$-MnPSe$_3$ and $i$-MnPSe$_3$ is highest around their respective band anticrossings (see Fig.~\ref{p4}). Therefore, the 2D materials that feature symmetry-protected Dirac and Weyl points and nodal lines may be used to designed AMs with strong anomalous Hall response.

\section{AHE\MakeLowercase{ in altermagnetic bilayer} M\MakeLowercase{n}S\MakeLowercase{e}}
\begin{figure}[H]
	\centering
	\includegraphics[width=17cm]{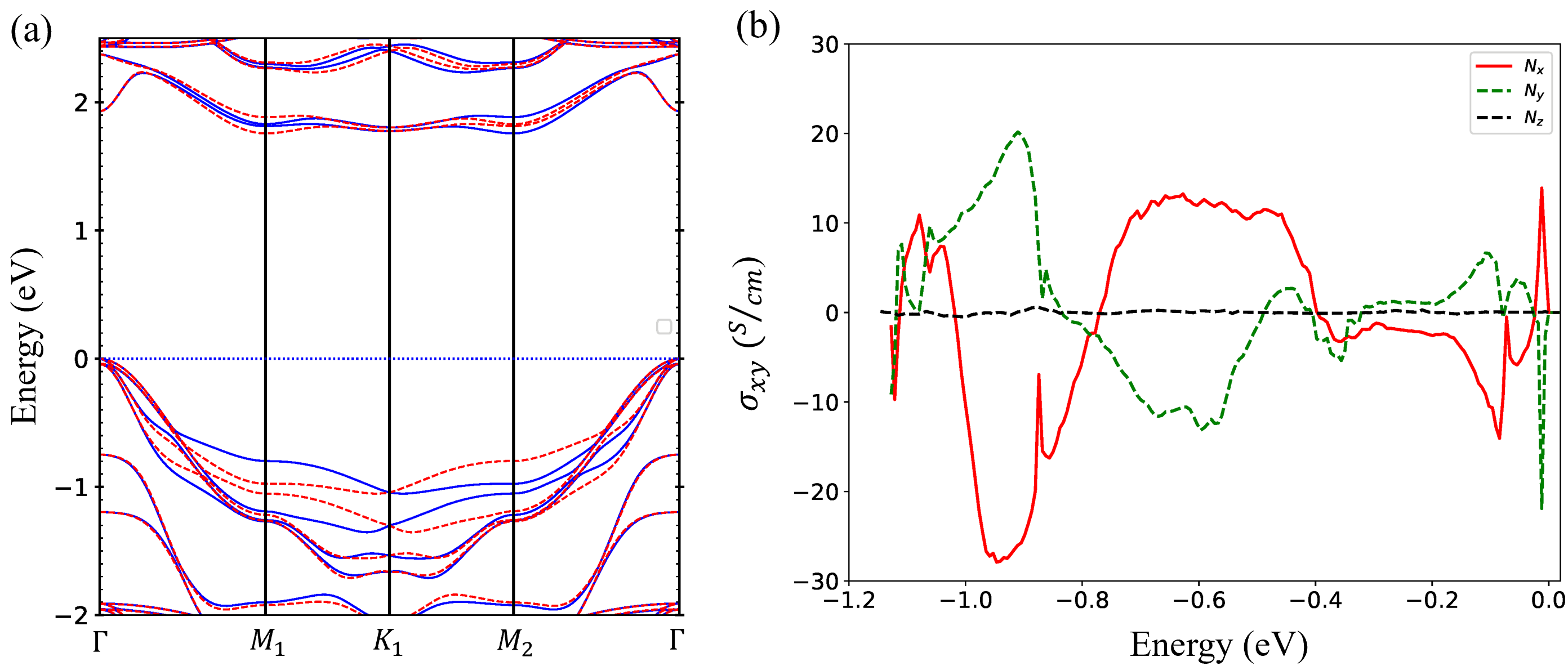}
	\caption{ (a) Spin polarized bands (without SOC) and (b) anomalous Hall conductivity $\sigma_{xy}$ of altermagnetic bilayer MnSe. The zero Fermi energy corresponds to the valence band maximum.}
	\label{p5}
\end{figure}

{To show that our group theoretic analysis is general and applies to other materials as well,} we also performed the calculations on bilayer MnSe. First we obtained the bilayer MnSe by taking the mirror reflection of the lower layer, followed by the in-plane translation of $1/3\boldsymbol{a}+1/3\boldsymbol{b}$. The spin polarized band structure of bilayer MnSe shows the $d$-wave altermagnetism [see Fig.~\ref{p5}(a)]. The opposite spin sublattices are connected by $C_2^{-\sqrt{3}x+y}$. The altermagnetic bilayer MnSe show anomalous Hall response similar to $d$-MnPSe$_3$ [see Fig.~\ref{p5}(b)].

\section{AHE \MakeLowercase{in weakly ferromagnetic bilayer} M\MakeLowercase{n}S\MakeLowercase{e}}
\begin{figure}[H]
	\centering
	\includegraphics[width=17cm]{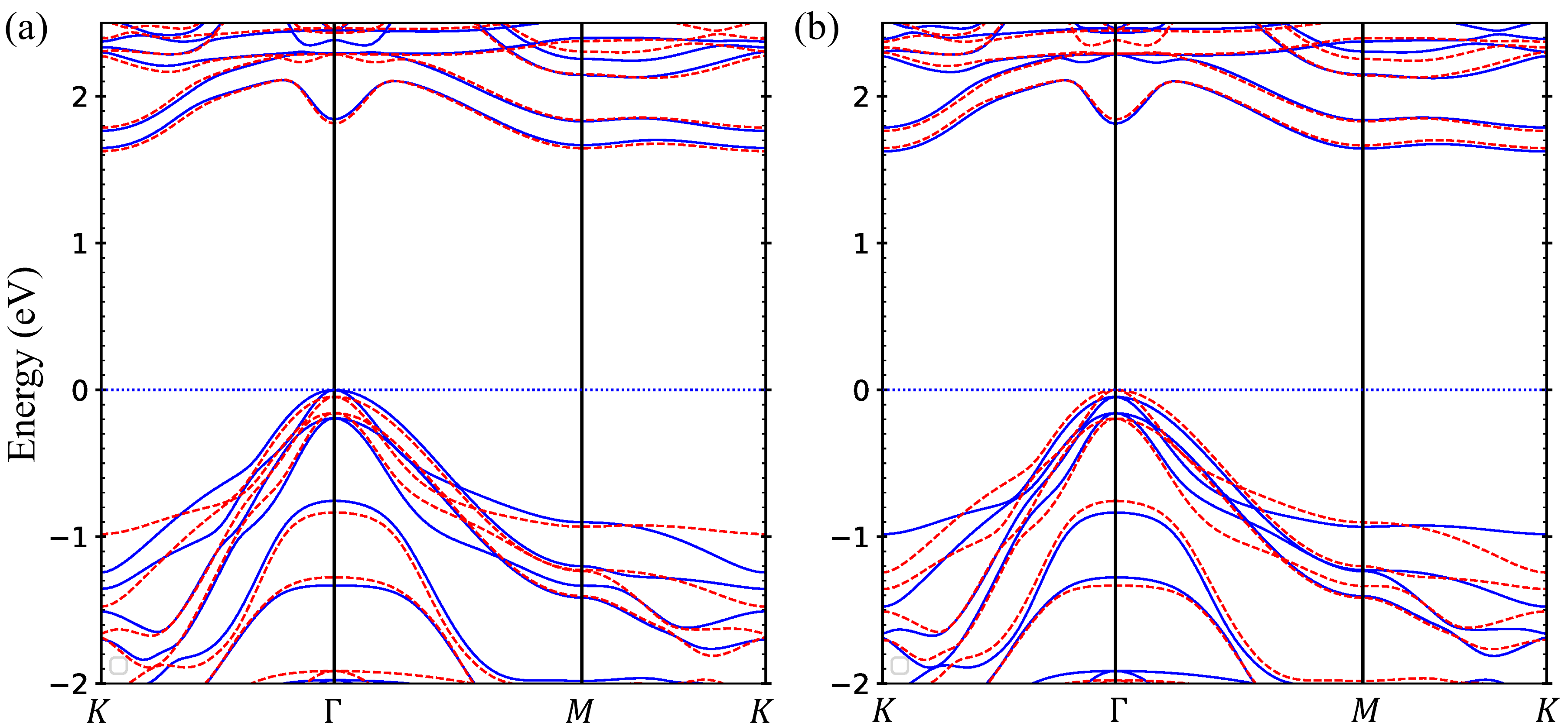}
	\caption{Spin polarized bands of (a) AB-MnSe  and (d) BA-MnSe without SOC. The zero Fermi energy corresponds to the valence band maximum.}
	\label{p6}
\end{figure}

\begin{figure}[H]
	\centering
	\includegraphics[width=17cm]{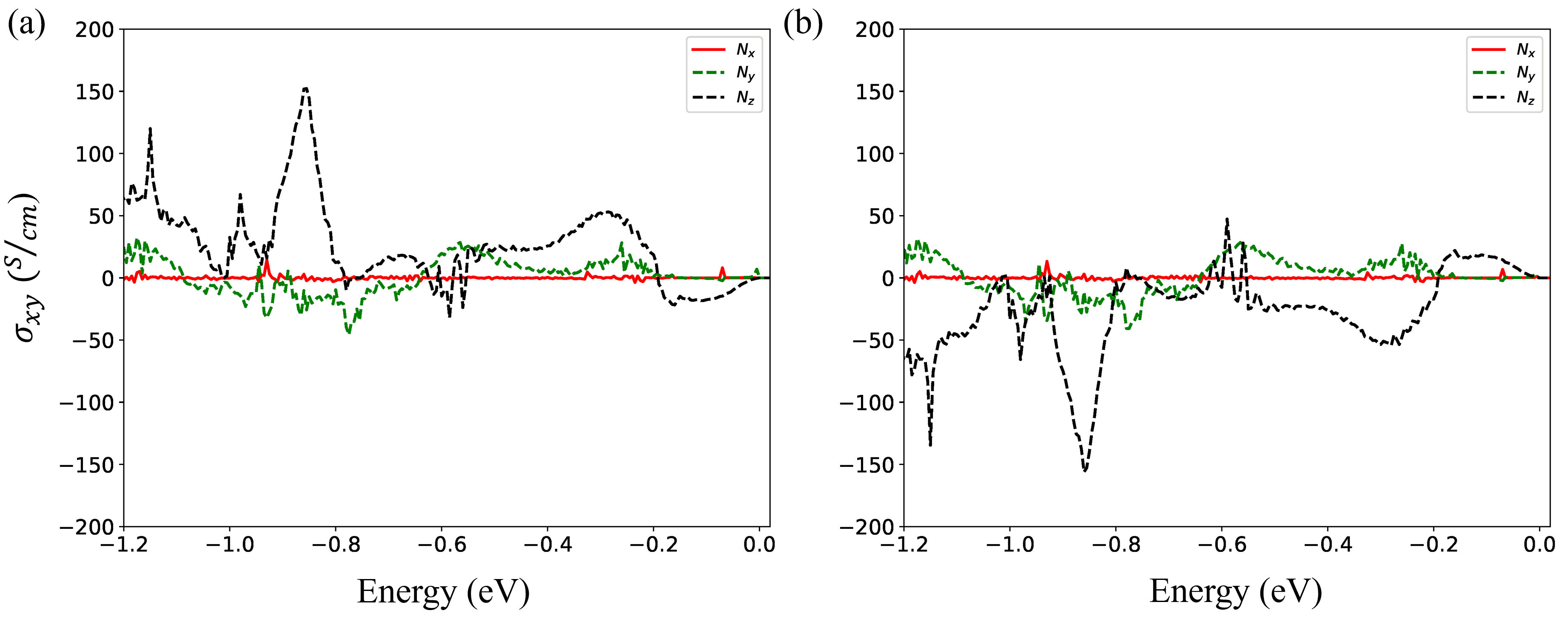}
	\caption{Anomalous Hall conductivity $\sigma_{xy}$ of (a) AB-MnSe  and (d) BA-MnSe as a function of Fermi energy for different orientations of the N\'eel vector. The zero Fermi energy corresponds to valence band maximum.}
	\label{p7}
\end{figure}

In this section, we discuss how AHE of an altermagnetic bilayer can be distinguished from the weakly ferromagnetic bilayer. For this, we used the bilayer MnSe obtained using the basal plane translation $1/3\boldsymbol{a}+2/3\boldsymbol{b}$ and $2/3\boldsymbol{a}+1/3\boldsymbol{b}$. Following Ref~\cite{liu2023tunable}, we name those stacking arrangements as AB-MnSe and BA-MnSe. The sliding ferroelectricty in bilayer MnSe leads to nonzero net magnetic moment (weak ferromagnetism). A recent study [$npj$ Computational Materials \textbf{9} (16), 2023] highlights the weak ferromagnetism in AB-MnSe and BA-MnSe, which is switchable by interlayer sliding from AB to BA stacking. We have plotted the bandstructure of AB-MnSe and BA-MnSe in Figs.~\ref{p6}(a) and ~\ref{p6}(b). The weak ferromagnetism can be understood from the fact that no crystal symmetry connects opposite spin sublattices in these structures. The weak switchable ferromagnetism is evident from the spin polarized band structure in Figs.~\ref{p7}(a) and ~\ref{p7}(b). Next, we obtained the AHC, $\sigma_{xy}$, of the weakly ferromagnetic bilayer MnSe for different N\'eel vector orientations. We observe that this weakly ferromagnetic bilayer MnSe shows FM-like Hall response where, the $\sigma_{xy}$ is nonzero when the N\'eel vector points along $z$-direction. {Note that $\sigma_{xy}$ is reversed not only upon switching the N\'{e}el vector from the $z$ to -$z$ direction, but also by changing the stacking from AB to BA.} These FM-like Hall responses in bilayer AFMs have recently gained much attention. For example, FM-type Hall responses have been reported for MnBi$_2$Te$_4$ through sliding ferroelectricity ~\cite{peng2023intrinsic,cao2023switchable} or external electric field~\cite{gao2021layer} 


\section*{References}